\documentclass[prd, twocolumn, nofootinbib, floatfix]{revtex4-1}

\usepackage{amsmath}
\usepackage{graphicx}
\usepackage{dcolumn}
\usepackage{bm}
\usepackage{enumerate} 
\usepackage{epsfig}
\usepackage{amssymb,latexsym,mathrsfs}
\usepackage{graphicx}
\usepackage{color}
\usepackage{hyperref}

\hypersetup{
    colorlinks=true,
    linkcolor=red,
    citecolor=blue,
} 

\newcommand{\be}{\begin{equation}}
\newcommand{\ee}{\end{equation}}
\newcommand{\bes}{\begin{equation*}}
\newcommand{\ees}{\end{equation*}}
\newcommand{\ra}[1]{\renewcommand{\arraystretch}{#1}}
\newcommand{\bs}{\begin{split}} 
\newcommand{\bea}{\begin{eqnarray}}
\newcommand{\eea}{\end{eqnarray}}

\newcommand{\lcdm}{$\Lambda$CDM } 
\newcommand{\mn}{{\mu\nu}} 
\newcommand{\ym}{{\rm YM}}

\begin{document}

\title{Cosmological consequences of classical flavor-space locked gauge field radiation} 
\author{Jannis Bielefeld and Robert R. Caldwell} 
\affiliation{Department of Physics \& Astronomy, Dartmouth College, Hanover, New Hampshire 03755 USA} 

\begin{abstract}
We propose a classical SU(2) gauge field in a flavor-space locked configuration as a species of radiation in the early Universe, and show that it would have a significant imprint on a primordial stochastic gravitational wave spectrum. In the flavor-space locked configuration, the electric and magnetic fields of each flavor are parallel and mutually orthogonal to other flavors, with isotropic and homogeneous stress energy. Due to the non-Abelian coupling, the gauge field breaks the symmetry between left- and right-circularly polarized gravitational waves. This broken chiral symmetry results in a unique signal: nonzero cross-correlation of the cosmic microwave background temperature and polarization, $TB$ and $EB$, both of which should be zero in the standard, chiral symmetric case.  We forecast the ability of current and future cosmic microwave background experiments to constrain this model. Furthermore, a wide range of behavior is shown to emerge, depending on the  gauge field coupling, abundance, and allocation into electric and magnetic field energy density. The fluctuation power of primordial gravitational waves oscillates back and forth into fluctuations of the gauge field. In certain cases, the gravitational wave spectrum is shown to be suppressed or amplified by up to an order of magnitude depending on the initial conditions of the gauge field.

%We propose a SU(2) Yang-Mills model of color radiation. We compute the effects of this fluid on a \lcdm cosmology. We analyze the full spectrum of scalar and tensor perturbations and show how vector modes decay. We explore an interesting property of gravitational waves: the YM fluid breaks the symmetry between left- and right handed gravitational waves. Physically, this is caused by the interplay of the fixed group space metric with the perturbed space-time metric. We introduce three new parameters that describe the amount of YM fluid, the initial conditions of the background solution, and the coupling constant of the YM fluid. We show that this YM fluid leaves a novel imprint on CMB anisotropies and gravitational wave spectra. We find that energy from primordial gravitational waves oscillates back and forth with the YM fluid which introduces features in the gravitational wave energy density spectrum larger than the standard model deviations. Significantly, we show that the YM fluid induces parity violating properties which we can use to test this model by computing $TB$ and $TE$ cross-correlations. The YM fluid has the potential to enhance the polarization autocorrelation spectrum by an order of magnitude. Finally, we compute how tightly future surveys constrain this model.
\end{abstract}

\date{\today} 

\maketitle
{\hypersetup{linkcolor=black}
	\tableofcontents
}

%%%%%%%%%%%%%%%%%%%%%%%%%%%%%%%%%%%%%%%%%%%%%%%%%%%%%%%%%%
\section{Introduction} 

The recent flourish of attention to the imprint of gravitational waves on the cosmic microwave background (CMB) \cite{Ade:2014xna, Ade:2015tva} has spawned an array of models that affect the physics of these tensor modes. On the observational side, the lensing $B$-mode spectrum of the CMB has been measured to ever increasing accuracy and $B$ mode surveys are getting close to detecting the primordial spectrum \cite{Smith:2007rg, Das:2011ak, Hanson:2013hsb, Ade:2014afa, Ade:2015tva, Keisler:2015jua, Ade:2015fwj}. This polarization information will  put stringent constraints on theories of the early Universe. However, the  subsequent propagation of tensor modes is also affected by the composition of the Universe which leaves an imprint on the tensor spectrum \cite{Bond:1996,Weinberg:2003,Pritchard:2004qp,Watanabe:2006qe}. The prospect of precision measurements of cosmological tensor modes opens up this area of research extending the era of precision cosmology to more intricate models concerning late-time tensor propagation.

A particular class of ideas proposes cosmological parity rotation (CPR). The standard model of particle physics is invariant under simultaneous reversals of charge (C), parity (P) and time (T). However, it does \emph{not} obey parity invariance on its own, most famously demonstrated for weak interactions \cite{Lee:1956qn}. This motivates similar behavior in cosmological physics and CPR is an example thereof. Among other symptoms CPR affects properties of gravitational waves as they possess intrinsic parity. First and foremost, allowing for different behavior for left- and right handed tensor modes results in CMB cross-correlations that are typically not allowed. Some mechanisms that cause this physics have their roots in novel interactions of electromagnetism, as for example cosmological birefringence \cite{Carroll:1998zi}, which rotates $E$ into $B$-modes. On the other hand the gravity sector itself can source the parity violation through explicit changes in the action. A wide variety of theories has been developed that accommodates these effects \cite{Lue:1998mq, Contaldi:2008yz, Takahashi:2009wc}.

Parity violation happens naturally in the weak sector of the standard model of particle physics. However, the standard model is presumably only a low-energy limit of a grand unified theory. In cosmology both inflation and late-time cosmic acceleration require new physics beyond general relativity -- and from a particle physics point of view -- beyond vanilla $\Lambda$CDM. It is therefore natural to ask whether the new physics responsible for cosmic inflation or late-time acceleration incorporates parity-violating effects. 

Inflation is usually thought of being caused by a scalar field. However, there has been a growing number of vector and gauge field models appearing in the literature over the past decades \cite{Caprini:2014mja, Anber:2006xt, Ford:1989me, Bento:1992wy, ArmendarizPicon:2004pm, Hosotani:1984wj, Galtsov:1991un, Kiselev:2004py, Carroll:2004ai, Boehmer:2007qa, Koivisto:2008xf, Bielefeld:2014nza}. In this paper we take inspiration from vector field inflation and Yang-Mills (YM) theories to introduce a fluid that causes parity-violating effects in the Universe. Field configurations that render the YM field homogeneous and isotropic have been found and these are employed to hold up isotropy in the Universe. We propose adding a dark YM radiation fluid to the energy budget of the Universe. This YM fluid violates parity on cosmological scales and changes the behavior of gravitational waves significantly. We use this as a simple model for chiral effects in gravity without requiring changing the Einstein-Hilbert action.

In this paper we will show the effect of the YM fluid on cosmological observables. We compute the change in gravitational wave amplitudes and their corresponding energy densities. We determine how this causes deviations from vanilla \lcdm in the CMB spectra and conclude by computing forecasted constraints that measurements of the classically vanishing cross-correlations impose.

The paper is structured as follows: We introduce the model in Sec.~\ref{sec:model} and derive the perturbation equations for the scalar, vector and tensor modes in Sec.~\ref{sec:perturbations}. We solve these equations in Sec.~\ref{sec:sols} where we separate off the treatment of the CMB into Sec.~\ref{sec:cmb}. In Sec.~\ref{sec:forecasts} we derive constraints that experiments put on this model and we conclude in Sec.~\ref{sec:concl}. Details about the implementation of this model in CAMB \cite{Lewis:1999bs} and solutions to the background equation of motion can be found in the appendices.

%%%%%%%%%%%%%%%%%%%%%%%%%%%%%%%%%%%%%%%%%%%%%%%%%%%%%%%%%%
\section{Background Model} 
\label{sec:model} 

We consider the standard cosmological model with the sole addition of a new gauge field as a toy model for cosmological chirality violation. This effect is not equivalent to parity violation in fundamental physics where some terms in the Lagrangian  violate parity explicitly. In spirit, the introduction of a particular configuration of the gauge field rather acts like the vacuum expectation value of a scalar field spontaneously breaking symmetry. Here, the broken symmetry is chirality: Left- and right handed gravitational waves will behave differently upon introducing this field. Additionally, we treat the gauge field as a classical, low-energy limit of a YM gauge field. The action of the theory is given by
\be
	S = \int d^4 x \sqrt{-g} \left(\frac{1}{2} M_P^2 R + {\cal L}_m - \frac{1}{4} F_{I\mu\nu}F^{I\mu\nu}\right)
	\label{eqn:model}
\ee
where Greek letters are used to represent space-time indices, lowercase Latin letters are spatial indices, and uppercase Latin letters $I$ are reserved for the gauge group indices. External matter is included in the Lagrangian density $\mathcal{L}_m$. The constant $M_P^{-2} = 8\pi G$ is the reduced Planck mass. The kinetic term for the gauge field is canonical and clearly inspired by high-energy physics, specifically Yang-Mills gauge theory. Symmetry breaking is a common feature of these theories and we use similar interactions to model chiral effects in cosmology. Therefore the field strength tensor $F^I_\mn$ is taken directly from Yang-Mills-like theories and it reads
\be
	F^{I}_\mn  \equiv  \partial_\mu A^{I}_\nu - \partial_\nu A^{I}_\mu -g_{\rm YM}\epsilon^{IJK}A_{J\mu} A_{K\nu}
	\label{eqn:fieldstrength}
\ee
where $g_\ym$ denotes the group coupling constant. In the internal space indices get raised and lowered with the flat metric $\delta^I_J$. 

The fluid equations for the gauge field follow from varying the action (\ref{eqn:model}) with respect to $A^I_\mu$:
\be
	\nabla^\mu F^I_\mn= g {\epsilon^{IJ}}_K A_J^{\mu} F^K_{\mu\nu}
	\label{eqn:fluid}
\ee
and the stress energy tensor for the gauge field is
\be
	T_{\mu\nu} = F_{I \mu\sigma} F_{\nu\tau}^I g^{\sigma \tau} - \frac{1}{4} g_{\mu\nu}F^J_{\sigma\tau}  F_J^{\sigma\tau}.
	\label{eqn:stress-energy}
\ee
Many authors have used gauge fields in cosmology -- for inflation as well as for late-time behavior \cite{Caprini:2014mja, Anber:2006xt, Ford:1989me, Bento:1992wy, ArmendarizPicon:2004pm, Hosotani:1984wj, Galtsov:1991un, Kiselev:2004py, Carroll:2004ai, Boehmer:2007qa, Koivisto:2008xf, Bielefeld:2014nza}. A common issue is that spin-1 fields have a preferred direction and therefore break isotropy. Nevertheless, there is a mechanism that provides isotropic stress energy \cite{Gal'tsov:2008km, ArmendarizPicon:2004pm} which relies on the homomorphism between the group of space rotations, O(3), and the internal group space SU(2). Making this choice for the gauge group, too, allows for isotropy: aligning the global part of the gauge SU(2) with the rotational SU(2) restores rotation symmetry because gauge fields are only defined up to these gauge transformations \cite{SheikhJabbari:2012qf}. Therefore $I \in \{1,2,3\}$ for each of the SU(2) generators.

We consider cosmological solutions wherein directions of the internal SU(2) space are aligned with the principle axes of the Cartesian, spatially flat Robertson-Walker space-time, $ds^2 = -dt^2 + a^2(t) d\vec x^2$. Specifically, we assume a \emph{flavor-space locked} configuration for the gauge field, wherein
\be
	A_\mu^I = \begin{cases} 
		\phi (t) \delta^I_i  & \mu = i \\
		0 & \mu = 0
  \end{cases}
  \label{eqn:assgm}
\ee
with all other components vanishing where the field $\phi(t)$ is a homogenous scalar. This particular configuration is just an example for the group of assignments of $\phi(t)$ to the gauge field that is spanned by all SU(2) gauge transformations on $A_\mu^I$. In particular one can rotate the $\phi(t)$ scalar inside $A_\mu^I$ without affecting the background or perturbation equations. This just follows from gauge-invariance of the YM Lagrangian under SU(2) transformations $A_\mu^I \to A_\mu^I + \epsilon^I_{\,JK}\Theta^J A_\mu^K + \frac{1}{g} \partial_\mu \Theta^I$ where $\Theta^I$ is a local, infinitesimal SU(2) gauge transformation. For the rnonzeroest of this analysis we choose the mapping in Eqn.(\ref{eqn:assgm}). 

With this assignment the nonzero components of the field strength tensor are
\bes
	F^I_{0i} = \dot\phi \delta^I_i\ \quad \text{and} \quad F^I_{ij} = -g_\ym \phi^2 {\epsilon^I}_{ij}
\ees
where a dot denotes a derivatives with respect to cosmic time. This field configuration resembles a condensate of massless gauge bosons, like the electroweak bosons before symmetry breaking, with their three hypercharge spin vectors pointing along the $x,\, y$ and $z$ directions. The Lagrangian expressed in terms of the scalar $\phi$ is
\be
	F^I_{\sigma\tau}  F_I^{\sigma\tau} = -6 \frac{\dot \phi^2}{a^2} + 6 {g_\ym}^2 \frac{\phi^4}{a^4} \>\> .
	\label{eqn:kin}
\ee
Using the above equations, from Eqn.~(\ref{eqn:stress-energy}) the energy density and pressure of the gauge field are
\be
	\begin{aligned}
		\rho &= \frac{3}{2}\left( \frac{\dot\phi^2}{a^2} + {g_\ym}^2 \frac{\phi^4}{a^4}\right) \cr
		p &= \frac{1}{2}\left( \frac{\dot\phi^2}{a^2} + {g_\ym}^2 \frac{\phi^4}{a^4}\right)
	\end{aligned}
	\label{eqn:rhop}
\ee
therefore, the standard, massless SU(2) gauge field has an equation of state of $w=1/3$, like radiation. The energy density in Eqn.~(\ref{eqn:rhop}) is composed of `electric' $\dot{\phi}$ and `magnetic' $g_\ym \phi^2$ contributions, motivated by the term structure of the gauge field. 

To gain some insight into the dynamics, we isolate the YM sector of the model. We write the gauge field action in the spatially flat Robertson-Walker coordinates and plug in the above expressions, Eqn.~(\ref{eqn:kin}), but then switch to conformal time whereby 
\begin{eqnarray*}
	S&=&\int dt \, d^3x\, \frac{3}{2}\left( a\dot{\phi}^2 - {g_\ym}^2 \frac{\phi^4}{a}\right)  \cr
	&=& \int d\tau \, d^3 x \frac{3}{2}\left( \phi'^2 - {g_\ym}^2 \phi^4\right),
\end{eqnarray*}
and derivatives with respect to conformal time are denoted with a prime. The equation of motion for this field in conformal time is
\be
	\phi'' + 2 {g_\ym}^2 \phi^3=0 .
	\label{egn:bgeom}
\ee
Interestingly, this description in terms of conformal time is independent of the scale factor, which allows for a general, analytic exploration of the properties of this field. Some details are given later in this section and Appendix~\ref{sec:app2}.

The solution may be expressed in the form of a Jacobi elliptic function \cite{ByrdFriedman}
\be
	g_\ym \phi(\tau) = c_1 \, {\rm sn}(c_1(\tau - \tau_i) +c_2 |-1).
	\label{eqn:analytic}
\ee
The constants are determined at the initial time $\tau_i$ as 
\be
	\begin{aligned}
		c_1^4 &= g_{\rm YM}^2(\phi_i'^2 + g_{\rm YM}^2 \phi_i^4) \\
		c_2 &= F( \csc^{-1}\left( 1 +  {\phi_i'^2}/{g_{\rm YM}^2 \phi_i^4}  \right)^{1/4} |-1),
	\end{aligned}
	\label{eqn:c1c2}
\ee 
where $F$ is an elliptic integral of the first kind. As shown in Appendix~\ref{sec:app2}, $\phi$ is an oscillating function of time. The smaller the value of $g_\ym$, the slower the rate of oscillation. However, for any coupling, the field behaves as radiation with equation of state $w=1/3$. 

We rephrase the constants $c_1$ and $c_2$ in terms of a parameter describing the energy density in the field, and a parameter describing the allocation of energy in the electric and magnetic components. The energy density in the gauge field as a  fraction of the photon energy density is $R_\ym = \rho_\ym / \rho_\gamma$ so that the initial values of the field and its derivative are
\be
	\begin{aligned}
		&\phi'_{ i} = H_0 M_P \sqrt{2 R_\ym \Omega_{\gamma, 0} \sin^2\theta} \\
		&\phi_{i}^2 =\frac{H_0 M_P}{g_\ym} \sqrt{2 R_\ym \Omega_{\gamma, 0} \cos^2\theta}
	\end{aligned}
	\label{eqn:initial}
 \ee
where $\theta \in[0,\frac{\pi}{2}]$ dials between electric and magnetic field energy. It will be useful to introduce two inverse comoving length scales
\be
\begin{aligned}
& k_0  \equiv \sqrt{2 \Omega_{\gamma,0} R_\ym} H_0 a_0 \\
& k_g \equiv \sqrt{g_\ym M_P a_0 k_0}
\end{aligned}
\label{eqn:k0gdef}
\ee
so that $c_1 = k_g$ and $c_2 = F(\csc^{-1}(\sqrt{\sec\theta})|-1)$ in Eqn.~(\ref{eqn:c1c2}). The gauge field oscillates with period 
\bes
	\tau = \frac{\Gamma(\tfrac{1}{4})^2}{\sqrt{2 \pi} k_g},
\ees
but its energy density and pressure scale with equation of state $w=1/3$, like radiation. For the field to remain coherent on cosmological time scales (and avoid a secular instability in the linear perturbations), and thereby have the maximum effect on cosmological physics, the coupling must be exponentially small,  $g_{\rm YM}\sim {\cal O}(H_0/M_P)\sim 10^{-60}$. This small number can be achieved if our theory originates with a dilatonlike factor $e^{\sigma} F^2$. With this coupling a dilaton can roll high up $e^\sigma$ without additional cost since $F^2$ vanishes under equipartition of energy between electric and magnetic modes. Supposing that a mechanism stabilizes $\sigma$ there would be no effect on the equations of motion, but the gravitation of the gauge field stress energy tensor would be magnified by this factor. Since our requirement that the field remains coherent on cosmological time scales is equivalent to setting $k_g \lesssim H_0$, then an original coupling $g \sim {\cal O}(1)$ can be engineered if a suitably large value of $\sigma$ is permitted. An origin for this small coupling as well as the flavor-space locked configuration might be devised in an inflationary epoch along the lines of \cite{Caldwell:2011ra, Motta:2012rn}, although that is beyond the scope of our present investigation. Inflationary scenarios based on a similar gauge field that include self-interactions with couplings to matter fields, have been studied elsewhere \cite{Maleknejad:2011jw,Adshead:2012kp}. 

Other potential constraints on this model could come from electric dipole measurements (EDM) assuming that baryons and leptons are charged under the gauge group of this field. Generally, EDMs violate both parity and time symmetries and they yield model-independent measures of CP violation in nature. However, as we will show now, the effects of the YM field on such a dipole relaxation process are negligible. We assume that a particle of mass $M$ possesses an intrinsic dipole moment $\mu = g_{\rm YM}/2M$ under this gauge group. The spin-flip energy due to the coupling between the dipole moment and the flavor-space locked field is therefore $2\mu B$ where $B = g_{\rm YM} \phi^2$. Using $B \sim M_P H_0 \sqrt{R_{\rm YM}/z_{eq}}$
where $z_{eq}$ is the redshift of radiation-matter equality, and $g_{\rm YM} \sim H_0/M_P$, then the energy shift is roughly $(H_0^2/M) \sqrt{R/z_{eq}}$. Using these results, we see that particles with mass $M$ produce photons with a wavelength $\sim H_0^{-1} (M/H_0)$, which for any reasonable mass is many  orders of magnitude longer than the Hubble horizon scale today.

%%%%%%%%%%%%%%%%%%%%%%%%%%%%%%%%%%%%%%%%%%%%%%%%%%%%%%%%%%
\section{Perturbation Equations} 
\label{sec:perturbations}

In this section we study the most general perturbations of the background solution presented in Sec.~\ref{sec:model}. We are interested in linear perturbations in this paper. These can be split up into scalar, vector and tensor perturbations that are decoupled from each other.

We are following Subsection III A of \cite{Dimastrogiovanni:2012ew} for developing the perturbation quantities. In general there are 9 perturbation degrees of freedom for the gauge field and 6 in the metric sector. Some of the perturbation quantities are defined as gradients of scalars, just as in general scalar-vector-tensor decompositions. To simplify notation and make some calculations easier, without loss of generality due to rotational invariance, the authors of \cite{Dimastrogiovanni:2012ew} choose the Fourier wave vectors $\vec{k}$ to be oriented along the $z$ direction. We follow that convention. Therefore, these perturbations appear solely in the $z$ direction and gradients turn into partial derivatives along $z$, $\partial_z$. We end up with the perturbations
\be
	\begin{aligned}
		\text{tensor:} \qquad &\delta A_{I \mu} = a \, t_{ij}\delta^i_I \delta^j_\mu \\ 
		&\delta g_{\mu\nu} = a^2 h_{\mu\nu} \\ \\
		\text{vector:} \qquad &\delta A^I_\mu = a\left( Y^I,0 ,0, \partial_z M^I \right) \\
		&\delta g_{0i}=a^2 B_i , \> \text{where } I, i \in \{1, 2\} \\ \\
		\text{scalar:} \qquad &\delta A_\mu^1 = a(0,\delta B,0,0) \\
		&\delta A_\mu^2 = a(0,0,\delta B,0)\\
		&\delta A_\mu^3 = a(\partial_z Y,0,0, \delta B + \partial^2_z M) \\
		&\delta g_{00} = a^2 2\Phi_G, \quad \delta g_{03} = a^2 \partial_z b 
	\end{aligned}
	\label{eqn:perts}
\ee
which show that there are five modes in the scalar sector, six in the vector sector, and four modes in the tensor sector. In the following subsections we will analyze these in detail. Not all of those will turn out to be physically propagating degrees of freedom. Some of them will have nondynamical, algebraic, or first-order time derivative equations of motion and we will refer to those equations as constraints and nondynamical modes.

In total, we will see that the theory in Eqn.~(\ref{eqn:model}) has 8 physical degrees of freedom -- 2 in the scalar, 2 in the vector, and all 4 in the tensor sector. The equations of motion for the perturbations follow from the fluid equation (\ref{eqn:fluid}). We will treat the three sectors individually in the next three subsections. 

%%%%%%%%%%%%%%%%%%%%%%%%%%%%%%%%%%%%%%%%%%%%%%%%%%%%%%%%%%
\subsection{Tensors}
\label{sec:tensperts}

The tensor perturbations describe gravitational wave propagation in this gauge fluid. The behavior of these is central to this discussion because the broken parity symmetry will become visible in this sector. Left- and right handed gravitational waves turn out to obey different equations of motion and therefore evolve differently from each other. Ultimately this has an imprint on the polarization of the CMB.

To obtain the tensor equations of motion we expand the action of Eqn.~(\ref{eqn:model}) up to second-order in the perturbation quantities and solve the corresponding Euler-Lagrange equations. Note that one has to explicitly add an external matter Lagrangian to obtain the correct equations as gravity couples to all stress energy. Without these the Friedmann equations would be sourced solely by the gauge field. The tensor perturbations in Eqns.~(\ref{eqn:perts}), $t_{ij}$ and $h_{ij}$, are transverse, traceless, synchronous tensors, following Refs.~\cite{Dimastrogiovanni:2012ew,Namba:2013kia}. Additionally, $t_{ij}$ and $h_{ij}$ are invariant under general covariance and SU(2) gauge transformations. Therefore, we adapt the usual $+$ and $\times$ polarization description of gravitational waves with
\bes
	\delta g_{\mu\nu} = a^2
	 \begin{pmatrix}
  		0 & 0 & 0 & 0 \\
  		0 & h_+ & h_\times & 0 \\
  		0  & h_\times  & -h_+ & 0  \\
  		0 & 0 & 0 & 0
 	\end{pmatrix}
\ees
which enter the tensor metric perturbations in Eqn.~(\ref{eqn:perts}). Next, we define left- and right handed circularly polarized wave amplitudes in the standard way,
\bes
	h_{L,R} = \frac{1}{\sqrt{2}}\left(h_+ \pm i h_\times\right), \qquad t_{L,R} = \frac{1}{\sqrt{2}}\left(t_+ \pm i t_\times\right),
\ees
upon which the left- and right handed systems separate.

Following through with the procedure outlined above we obtain the tensor perturbation propagation equations: the equations of motion for the Fourier amplitudes of a right-circularly polarized gravitational wave traveling in the $+z$ direction, and the corresponding gauge field fluctuation, are given by
\begin{widetext}
\begin{eqnarray}
&&h_R''+ 2 \frac{a'}{a}h_R' +\left[k^2  + \frac{2}{a^2 M_P^2}\left(g_{\rm YM}^2 \phi^4 - \phi'^2\right)\right] h_R =  
 	-\frac{2}{a M_P}\left[ (k - g_{\rm YM} \phi) g_{\rm YM} \phi^2 t_R + \frac{a'}{a}\phi' t_R + \phi' t'_R\right], \nonumber \\
&&t_R''+2 \frac{a'}{a}t_R' + \left[k^2 + \frac{a''}{a} - 2 k g_{\rm YM} \phi \right] t_R  =  - \frac{2}{a M_P}\left[(k+g_{\rm YM} \phi) g_{\rm YM} \phi^2 h_R - \phi' h_R'\right]. 
	\label{eqn:Htens}
\end{eqnarray}
\end{widetext}
The left-circularly polarized gravitational wave propagation equations are obtained upon the exchange $k \to -k$. This has consequences for the evolution --- circular dichroism --- as can be seen by examining the terms with a single power of $k$ in the above equations. These terms show that the effective mass term $-2 g_{\rm YM} \phi$ for the YM tensor perturbation $t$ and the coupling between $h$ and $t$ differs for left and right circular polarizations.  For clarity, we have omitted the anisotropic shear contributed by other species, such as photons and neutrinos, although these effects are included in our CMB analysis. 

Examining the form of the gravitational wave equation in the presence of the gauge field, we make the following observations. First, the parity violation depends solely on the coupling $g_\ym$ and is due to the antisymmetric Levi-Civita tensor in the internal SU(2) space, which we have identified with the principle axes of our physical space. Therefore, the coupling constant $g_\ym$ will have a big impact on the chiral properties of this model. Second, the $t_{R/L}$ terms in the gravitational wave equations may be thought of collectively as representing an anisotropic shear source. This is similar to the sources arising from photons and neutrinos which damp subhorizon scale gravitational waves in the standard cosmological model. However, here they introduce the parity breaking features into cosmology. And finally, we note that the background gauge field contributes a novel masslike term involving differences in the background gauge electric and magnetic fields $ \propto( g_\ym^2 \phi^4 - \phi'^2)$. These equations will be used later in Sec.~\ref{sec:sols} to describe gravitational wave propagation and CMB temperature and polarization anisotropy spectra.

%%%%%%%%%%%%%%%%%%%%%%%%%%%%%%%%%%%%%%%%%%%%%%%%%%%%%%%%%%
\subsection{Vectors}
\label{sec:vec}

In this subsection we will derive the vector perturbation equations; we will show that they lead to subdominant effects in Sec.~\ref{subsec:vector}. Therefore, for the CMB results, we will be able to ignore them in our CAMB implementation.

The vector perturbations  of the metric and the gauge field are defined in Eqn.~(\ref{eqn:perts}). As it will turn out only $M^I$ is a dynamical field. To obtain the equations of motion we expand the Lagrangian to second-order in the vector perturbations, add an external matter part, which we choose to be massive scalar, and compute the Euler-Lagrange equations for every component of the vector perturbation quantities $\{Y_I, B_I, M_I\}$. This yields an equation for each component of the vectors of which there are two. These are
\begin{widetext}
\be
	\begin{aligned}
		Y_1 : \,\, & k^2(a M_1)' - i k g_\ym (\phi (a M_2)' - \phi' a M_2) - a(k^2 + 2 g_\ym^2\phi^2)Y_1 + 2 i a k g_\ym \phi Y_2 =  g_\ym \phi^2 (i k B_2 - 2 g_\ym \phi B_1) \\ \\
		B_1 : \,\, & M_P^2 (6 a'^2 - a^2 k^2) B_1 = (3 \phi'^2 - g_\ym^2\phi^4)B_1 + 2 a g_\ym \phi^2 (2 g_\ym \phi Y_1 - i k Y_2)+ 2 i g_\ym k \phi (\phi (a M_2)' - \phi' a M_2) \\ \\
		M_1 : \,\, & i k(a M_1'' + 2 a' M_1' + a'' M_1 + a g_\ym^2 \phi^2 M_1) = i k (a Y_1)' + g_\ym(\phi (a Y_2)' + 2 \phi' a Y_2) - g_\ym \phi(\phi B_2' + 3 \phi' B_2) \\ \\
	\end{aligned}
	\label{eqn:vecpert}
\ee
\end{widetext}
and the other set of equations is obtained by assigning $\{1,2, i\} \rightarrow \{2, 1, -i\}$. Conveniently the $Y_I$ equations can be solved algebraically and eliminated. The $B_I$ equations, the vector metric perturbations, also couple to any other source of vector modes such as the photon fluid. For simplicity, we have ignored external sources of vector perturbations in deriving the above, middle equation.

%%%%%%%%%%%%%%%%%%%%%%%%%%%%%%%%%%%%%%%%%%%%%%%%%%%%%%%%%%
\subsection{Scalars}

The scalar perturbations do not affect the chiral symmetry breaking in the theory, but to properly compute the impact of this theory on the CMB we need the propagation equations for them. Here we use the definitions in Eqn.~(\ref{eqn:perts}) together with energy conservation of the stress energy tensor in Eqn.~(\ref{eqn:stress-energy}) $\nabla_\mu {T^\mu}_\nu = 0$ and the fluid Eqns.~(\ref{eqn:fluid}) to obtain the full set of perturbation equations.
Defining the transformations
\begin{equation*}
	y=a Y , \hspace{1cm} \delta m = a k^2 M , \hspace{1cm} \delta \phi = a \delta B
\end{equation*}
will make these equations simpler. Using energy conservation and the fluid equations, the differential equations for the 4 new scalar quantities $\delta \phi$, $y$, $\delta m$ and $b$ read
\be
	\begin{aligned}
		&\delta m''-k^2 y' + k^2 \delta \phi + k^2 \phi' b = 0 \\
		&\delta \phi'' + k^2 \delta \phi -2{g_\ym}^2 \phi^2\left(\delta m -3\delta \phi\right)+k^2 \phi' b \\
		&\qquad - 4{g_\ym}^2 \phi^3 \Phi_G + \phi' \Phi_G' = 0
	\end{aligned}
	\label{eqn:scalphys}
\ee
together with the first-order constraint equations
\be
	\begin{aligned}
		&\delta m' - \delta \phi' - \left(k^2+2{g_\ym}^2 \phi^2 \right)y + 2{g_\ym}^2 \phi^3 b - \phi' \Phi_G = 0 \\
		&\phi\left(\delta m- 3\delta\phi\right)+\phi y' + 2 \phi' y + \phi^2 \Phi_G - 3 \phi' \phi \, b - \phi^2 b' = 0
	\end{aligned}
	\label{eqn:scalconstr}
\ee
The first set of equations (\ref{eqn:scalphys}) describes true physically propagating degrees of freedom for $\delta m$ and $\delta \phi$. The second set (Eqns.~(\ref{eqn:scalconstr})), however, are merely constraint equations that do not describe physical degrees of freedom ($y$, $b$). They lack second derivatives in a timelike component. 

We can also express these in more common fluid variables to shed some light on these quantities. Equations (21) and (22) in \cite{Ma:1995ey} define the energy density perturbation $\delta \rho$, the divergence of the fluid velocity $\theta$, and the shear stress $\sigma$. Reading these off from the perturbed stress energy tensor yields
\be
	\begin{aligned}
		&\rho \sigma = \frac{k^2}{a^4} \phi' y + {g_\ym}^2 \frac{\phi^3}{a^4}\delta m - \frac{\phi'}{a^4} \delta m' \\
		&(\rho + p)\theta = 2k^2 \frac{\phi'}{a^4}\delta\phi + 2{g_\ym}^2 k^2 \frac{\phi^3}{a^4}y - 2{g_\ym}^2 k^2 \frac{\phi^4}{a^4} b \\
		&\delta \rho = \frac{\phi'}{a^4}\left( 3\delta\phi' - \delta m'\right) + 2{g_\ym}^2 \frac{\phi^3}{a^4}\left(3\delta\phi - \delta m\right) \\
		&\qquad + 3\frac{{\phi'}^2}{a^4}\Phi_G + k^2 \frac{\phi'}{a^4}y \> .
	\end{aligned}
	\label{eqn:fluidvars}
\ee
The first of these equations is the most interesting one as it describes the scalar contribution to the shear. The set of equations that describe the evolution of the fluid variables is equivalent to Eqns.~(\ref{eqn:scalphys}, \ref{eqn:scalconstr}) and is given by
\be
	\begin{aligned}
		&\delta' = \frac{a'}{a}\delta -3\frac{a'}{a}\frac{\delta p}{\rho} - \frac{4}{3}\theta - \frac{4}{3}k^2 b \\
		&\theta' = k^2 \left(\frac{3}{4}\frac{\delta p}{\rho} - \Phi_G - \sigma \right) \\
		&\delta p = \frac{1}{3}\delta \rho .
	\end{aligned}
	\label{eqn:fluideom}
\ee
Combining these gives second-order equations again. The relationship between pressure and energy density perturbations is the same as for radiation. The scalar sector of the YM fluid behaves very similarly to regular radiation. This can be seen most easily by transforming into the conformal-Newtonian gauge using the usual transformation laws
\be
	\begin{aligned}
		&\frac{a'}{a}b = \Psi_{\rm gi} = -\phi_{cN} \\
		& b' +\frac{a'}{a}b = \Phi_{\rm gi} = \psi_{cN}
	\end{aligned}
	\label{eqn:gaugetrans}
\ee
where $\Psi_{\rm gi}$ and $\Phi_{\rm gi}$ are the gauge-invariant scalar potentials from \cite{Mukhanov:1990me}. The subscript $cN$ stands for the conformal Newtonian gauge from \cite{Ma:1995ey}, which is the gauge that we want to translate the perturbations into. With these we get
\be
\begin{aligned}
	(cN) \quad & \delta' = -\frac{4}{3}\theta + 4 \phi'_{cN}  \\
	(cN) \quad & \theta' = \frac{1}{4}k^2\delta + k^2 \psi - k^2\sigma ,
\end{aligned}
\ee
where the notation ``cN" is to remind us that all variables are now in the conformal-Newtonian gauge. This is identical to Eqns.~(64) in \cite{Ma:1995ey}. It is just the contribution to $\sigma$ in Eqn.~(\ref{eqn:fluidvars}) that gives deviations from normal radiation in the scalar sector. 

In the case $g_\ym =0$, the evolution of the scalar shear, 
\be
(cN) \quad \sigma' = \frac{2}{3} \theta,
\ee
may be obtained from the  perturbation equations. In the case $g_\ym \neq 0$, special care must be given to evolve the scalar shear as will be explained in Appendix~\ref{appendix}. To implement the code in CAMB we follow the definitions of the perfect fluid perturbations from \cite{Ma:1995ey} in the synchronous gauge. We keep using adiabatic initial conditions for the density contrast $\delta$ upon introducing the YM fluid.

%%%%%%%%%%%%%%%%%%%%%%%%%%%%%%%%%%%%%%%%%%%%%%%%%%%%%%%%%%
\section{Solutions of the Perturbation Equations}
\label{sec:sols}

With the evolution equations derived in the previous section, we can analyze the modified behavior of gravitational waves as well as their impact on the CMB. In the following sections we will study the cosmological evolution of gravitational waves and present a technique to efficiently obtain those results. Finally, we will determine the impact the YM fluid has on the power spectra.

%%%%%%%%%%%%%%%%%%%%%%%%%%%%%%%%%%%%%%%%%%%%%%%%%%%%%%%%%%
\subsection{Prologue: Secular perturbations}
\label{sec:secinst}

In the following section we will derive the scalar perturbation equations and compare these to the standard $\Lambda$CDM results in the conformal-Newtonian gauge. Before we make the formal derivation we will briefly review secular instabilities as an example how perturbations can grow indefinitely without actually affecting the physics.

Consider a pendulum of length $l$ swinging at an angle $\theta$ relative to the vertical. The full ordinary differential equation describing this system is
\be
   \ddot \theta + \kappa \sin\theta = 0
\ee
where $\kappa = g/l$. Note that expanding this equation up to third order in $\theta$ yields an equation similar to the equation of motion for our gauge scalar (see Eqn.~(\ref{egn:bgeom})). Introducing a perturbation $\theta = \theta_0 + \delta \theta$ yields
\be
   \ddot{\delta\theta} + \kappa \cos\theta_0 \delta\theta = 0 \>,
\ee
where $\theta_0$ follows the background trajectory. This is the linearized perturbation equation for this system. Now, as $\cos\theta_0$ oscillates, the perturbation $\delta\theta$ grows linearly which looks like an instability of the theory. This is traditionally called a secular perturbation. However, obviously the simple pendulum does not contain any unstable fluctuation. The actual problem lies in misinterpreting initial conditions for this perturbation \cite{Goldstein}. Shifting the initial conditions of the pendulum propagation might look like this growing mode. Secular perturbations often refer to these shifts in initial conditions and are therefore not a true physical instability.

To be explicit, let us consider a small perturbation to the gauge field solution (\ref{eqn:analytic}), $\phi \to \phi + \delta\phi$, and linearize the equation of motion (\ref{egn:bgeom}), whereby
\be
\delta\phi'' + 6g_\ym^2 \phi^2 \delta\phi = 0.
\label{eqn:linsec}
\ee
The solution to the above equation is
\be
\delta\phi = \frac{g_\ym}{c_1^2} \left( \delta\phi(\tau_i) \phi' + \frac{1}{2}\delta\phi'(\tau_i) (\tau \phi' + \phi)\right).
\label{eqn:linsecsol}
\ee
Although $\delta\phi$ contains a term that grows linearly in $\tau$, it is obvious that this signals no physical instability; instead, it indicates a limit to our linearization approximation. If we simply adjust the initial conditions for $\phi$ or else refrain from linearizing the perturbation equation, then we can make $\delta\phi$ well behaved.

We will encounter these explicitly in the solutions to the vector perturbations in Sec.~\ref{subsec:vector} and scalar perturbations in Sec.~\ref{sec:cmb}.

%%%%%%%%%%%%%%%%%%%%%%%%%%%%%%%%%%%%%%%%%%%%%%%%%%%%%%%%%%
\subsection{Gravitational Waves}
\label{sec:gw}

In this chapter we will use the formal results from Sec.~\ref{sec:tensperts} to describe how gravitational waves behave differently with the YM fluid present compared to standard $\Lambda$CDM. Most prominently the fluid introduces chiral effects into the tensor sector of gravity and therefore affects the behavior of left- vs right-handed polarized gravitational waves. This cosmological parity rotation is manifest in the coupling to gravitational waves. To visualize this, consider a gravitational wave passing through the gauge field described in the above sections. Generally speaking, gravitational waves  induce a quadrupolar distortion, alternately squeezing and stretching the stress and energy of the YM field. However, the gauge field itself possesses a preferred handedness via the right handed SU(2) structure constants. Since the fields are flavor-space locked, where the principal axes of SU(2) are aligned with the spatial coordinate basis, the gauge field stress energy will vibrate in sympathy to a right handed wave, and with antipathy to a left handed wave. A nice visualization for a similar effect is the rattleback top \cite{rattleback}. 

We now analyze the evolution of gravitational waves described in Sec.~\ref{sec:perturbations}. There is a rich variety of behavior in the evolution of this system, dependent upon the coupling $g_{\rm YM}$, the abundance $R_{\rm YM} = \rho_{\rm YM}/\rho_{\rm rad}$ during the radiation era, the relative contributions of electric and magnetic field energy, $\phi'$ and $g_{\rm YM} \phi^2$, and the initial conditions for the perturbations $h_{R/L},\, t_{R/L}$. For these purposes, we omit the anisotropic shear contributed by other species, such as photons and neutrinos, although these effects are included in our CMB analysis. Furthermore we assume a standard scale-free primordial tensor power spectrum.

%%%%%%%%%%%%%%%%%%%%%%%%%%%%%%%%%%%%%%%%%%%%%%%%%%%%%%%%%%
\subsubsection{Long wavelengths}
\label{sec:longwave}
 
We first tackle the long wavelength behavior of the gravitational waves. For simplicity, we begin our investigation with the simpler scenario, setting $g_\ym=0$ which corresponds to electrodynamics with three flavors or colors, also known as ``color electrodynamics". In this case, the distinction between $L-$ and $R-$handed gravitational waves is gone, so we label the tensor perturbations with the subscript ``$A$" for ambidextrous. The equations for $h_A$ and $t_A$ (\ref{eqn:Htens}) may be combined as
\begin{eqnarray}
&& h_A'' + 2 \frac{a'}{a}h_A' + (k^2 - 2 k_0^2\frac{a_0^2}{a^2})h_A \cr
&&\quad = -4 k_0^2\frac{a_0^2}{a^2} \int^\tau_{\tau_i} d\tau' \, h_A' \, \cos k(\tau-\tau').
\label{eqn:colorh}
\end{eqnarray}
Hence, there is a time-dependent effective mass term, $m_{\rm eff}^2 = k^2 - 2 k_0^2 a_0^2/a^2$.  In the case $k < \sqrt{2} k_0 a_0/a$ this term is negative, producing an enhancement of the gravitational wave amplitude. To see this, we take the long wavelength limit, $k\to 0$, in which case the equation becomes
\begin{equation}
h_A'' + 2 \frac{a'}{a}h_A'   +  2 k_0^2\frac{a_0^2}{a^2}h_A   = 4 k_0^2\frac{a_0^2}{a^2} h_A(\tau_i)
\end{equation}
where $h_A(\tau_i)$ is the initial value of the wave amplitude. By inspection we note that there is a fixed point where the amplitude approaches a constant value, $h_A \to  2 h_A(\tau_i)$. We show this more rigorously by assuming a radiation-dominated expansion scale factor $a=a_i  \tau/\tau_i$ and, assuming the standard initial conditions $h'_A(\tau_i)=0$, then
 \begin{eqnarray}
h_A(\tau) &=&  h_A(\tau_i) \left[ 2 + c_1 \left(\frac{\tau}{\tau_i}\right)^{n_1}  + c_2  \left(\frac{\tau}{\tau_i}\right)^{n_2} \right] \\
n_{1,2} &=& -\frac{1}{2}\left(1 \pm \sqrt{1 - 8 (a_0 k_0 \tau_i/a_i)^2}\right)
\end{eqnarray}
where $c_i = -(1+n_i)/(1 + 2 n_i)$. Since the exponents $n_{1,2}$ are both negative, the time-dependent terms decay, and the fixed point is soon reached. It is clear that for modes with wave number $k \tau_0 \ll 1$ and also $k \ll \sqrt{2} k_0 a_0/a$, the amplitude of long wavelength gravitational waves is doubled.  

We can extend the analysis of the behavior of long wavelength gravitational waves to the coupled, YM case. To begin, we set $k\to 0$, and assume a pure radiation expansion rate so that the $a''/a$ terms may be neglected. Next, we introduce a change of variable $u = a t$ and introduce the new time variable $x = k_g (\tau-\tau_i) + c_2$. The system of equations (\ref{eqn:Htens}) now becomes
\begin{eqnarray}
&&\frac{d^2 h}{dx^2} + 2 \frac{d \ln a}{dx}\frac{dh}{dx} + \frac{2}{a^2 \lambda^2}\left(\psi^4 - d\psi^2\right) h \cr\cr
&&\qquad = -\frac{2}{a^2\lambda}\left(d\psi \frac{du}{dx} + \frac{1}{2}\frac{d^2\psi}{dx^2} u\right)
\label{eqn:k01}
\\\cr
&&\frac{d^2 u}{dx^2} = \frac{2}{\lambda^2}\left(d\psi \frac{dh}{dx} + \frac{1}{2}\frac{d^2 \psi}{dx^2} h\right)
\label{eqn:k02}
\end{eqnarray}
where $\lambda = g_\ym a_0 M_P /k_g$. The functions $\psi$ and $d\psi$ are the background field and derivative scaled to unit amplitude as
\begin{equation}
\begin{aligned}
	\psi &= \frac{g_\ym}{k_g} \phi = {\rm sn}(x | -1) \\
	d\psi &= \frac{g_\ym}{k_g^2}\phi' = {\rm cn}(x | -1) \, {\rm dn}(x | -1).
\end{aligned}
\label{eqn:psi}
\end{equation}
In the case that $\psi,\,d\psi$ are slowly evolving, effectively constant, then we note there is a self-consistent fixed point solution whereby $h \to h_f$ and $u \to a t_f$ such that $h_f'$ and $t_f'$ are negligible. From Eqn.~(\ref{eqn:k01}) we determine
\begin{equation}
d\psi \frac{d(a t_f)}{dx} = -\frac{1}{\lambda}\left(\psi^4 - d\psi^2\right) h_f.
\end{equation}
Next, integrating Eqn.~(\ref{eqn:k02}) so that
\begin{equation}
\frac{du}{dx} = \frac{2}{\lambda}\left( \big[ d\psi\, h \big]_i^f - \frac{1}{2} \int dx \, \frac{d^2 \psi}{dx^2} \, h \right)
\end{equation}
and inserting it into the first equation, then we find at the fixed point
\begin{equation}
h(\tau_f) = \frac{2 d\psi^2}{\psi^4 + d\psi^2} h(\tau_i).
\end{equation}
These results are confirmed by a numerical integration. To reformulate these results in terms of the initial conditions for $\psi$ and $d\psi$, we obtain
\be
	\begin{aligned}
		h(\tau_f) &= 2 \sin^2\theta \, h(\tau_i) \\
		t(\tau_f) &= -2 (1+z_f) k_0 \tau_f \sin\theta \, \cos 2\theta \, h(\tau_i)
	\end{aligned}
	\label{eqn:transfer}
\ee
where $\tau_f,\,z_f$ are the conformal time and redshift and $\theta$ is defined in Eqn.~(\ref{eqn:initial}). These assume standard initial conditions for $h$ and we have set the initial perturbation in the YM fluid $t(\tau_i) = 0$. In the case $\theta = \pi/4$, the effective mass term in the evolution equation for $h$, proportional to the difference  between the background field electric and magnetic energy densities, vanishes, so that $h$ and $t$ do not evolve on long wavelengths. This is the standard case, which we focus on in this paper. 

In the case $\theta = \pi/2$, the $\psi^4$ term drops out of the effective mass term, just as in the case of color electrodynamics where $g_\ym=0$, and the amplitude of $h$ doubles. The most surprising case, however, seems to be when $\theta = 0$, so that the effective mass is entirely due to the $\psi^4$ term, which acts to damp the gravitational wave amplitude. Assuming that the evolution begins at some time shortly after inflation, then the fixed point is soon reached well in advance of horizon entry by any wavelengths of interest. The prediction, borne out by numerical integration, is that the gravitational wave and tensor modes of the gauge field are damped out. (We note that such suppression could modify the predictions of the gauge-flation and chromo-natural inflation scenarios \cite{Maleknejad:2011jw,Adshead:2012kp,Namba:2013kia} or other inflationary models that traditionally overproduce gravitational waves. This effect has not been explored before and might mitigate the shortcomings of these theories that ultimately led to the predictions that ruled them out.) 

This solution (Eqn.~(\ref{eqn:transfer})) also allows us to predict the amplitude of superhorizon modes at some time late in the radiation-dominated era, which we may then use as the initial data for a numerical study as the modes proceed to enter the horizon.

\begin{figure}[h]
\includegraphics[width=1\linewidth]{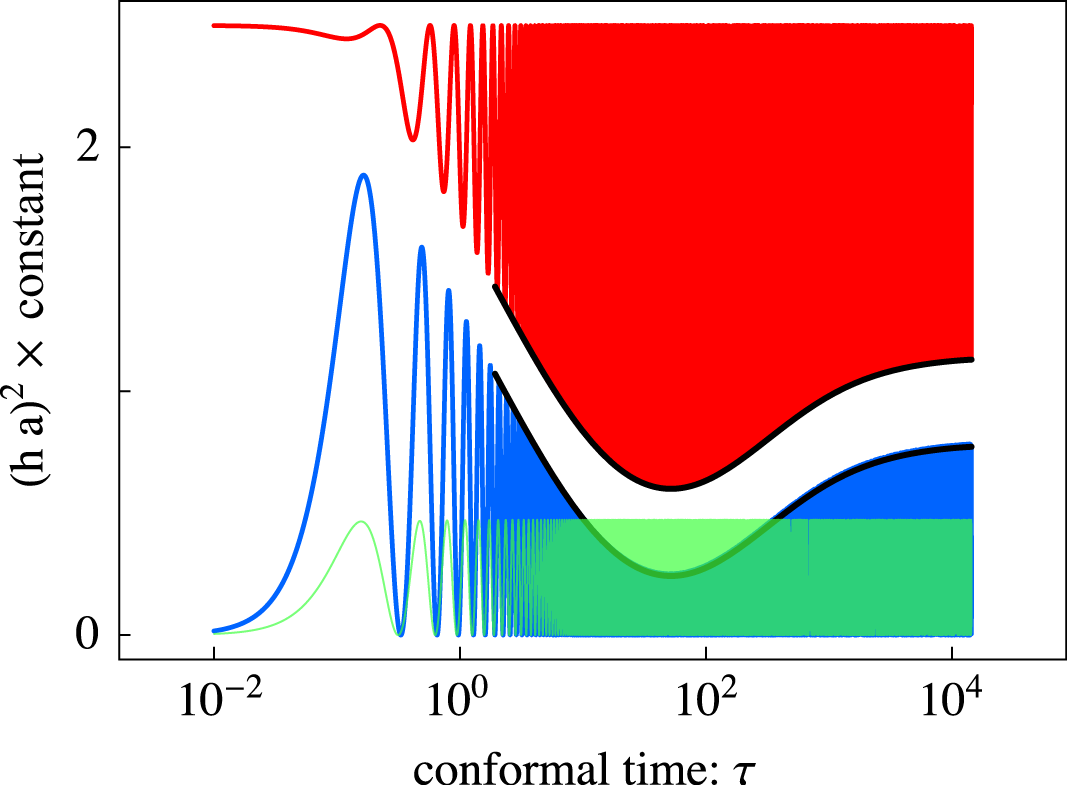}
\caption{Gravitational wave amplitude evolution as a function of conformal time is shown for the case $g_{\rm YM}=0$, $R_{\rm YM}=0.03$ (blue) and wave number  $k=10$~h/Mpc, as compared to the standard case $R_{\rm YM}=0$ (green). The excitations of the gauge field are shown (red) as an offset minus a constant times $(a t_{A})^2$, to illustrate their complementary behavior. The solid (black) lines show the results of WKB solutions for the envelopes of the oscillatory waveforms.}
\label{fig:fig1a}
\end{figure}

\begin{figure}[h]
\includegraphics[width=1\linewidth]{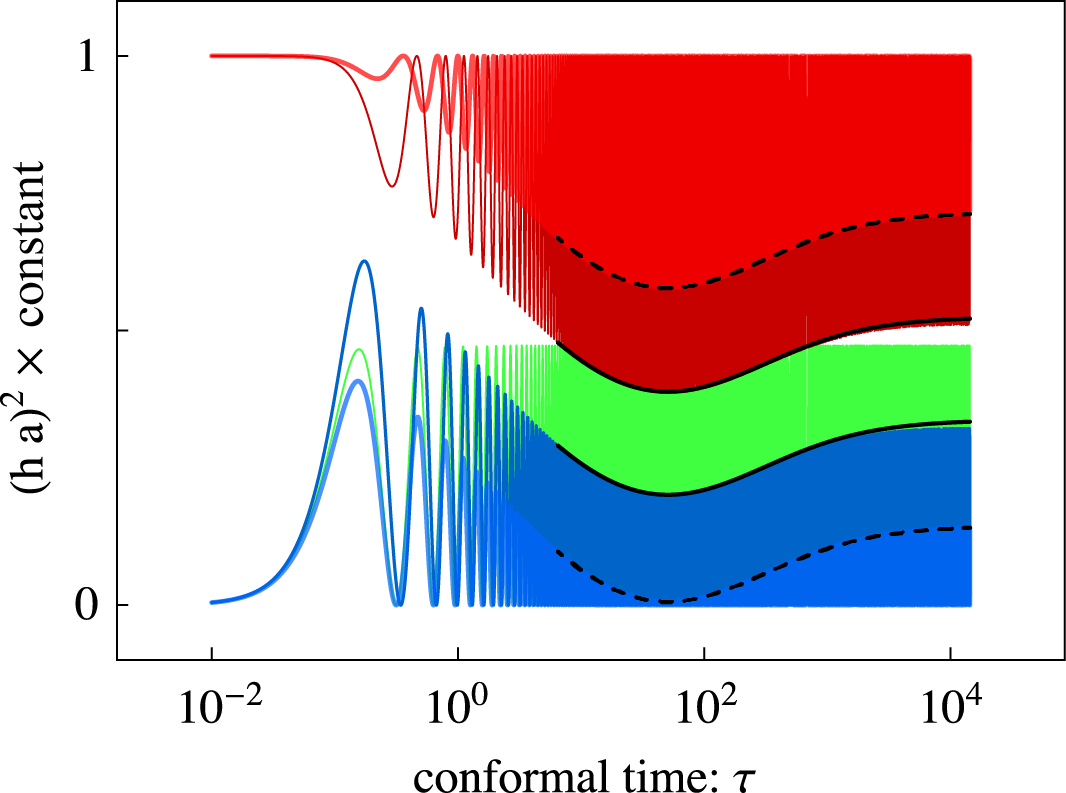}
\caption{As in Fig.~\ref{fig:fig1a}, the gravitational wave amplitude evolution as a function of conformal time is shown, this time for the case $g_{\rm YM}=10^{-60}$, $R_{\rm YM}=0.03$ (blue) and wave number  $k=10$~h/Mpc. The waveforms capped with solid lines are right handed; those with dashed lines are left handed.}
\label{fig:fig1b}
\end{figure}

We now focus on a minimal scenario in which the initial field energy of the YM fluid is split equally between the electric and magnetic field, $\phi' = g_{\rm YM} \phi^2$, whereby $\theta = \pi/4$. These initial conditions live closest to the standard cosmological model, as long wavelength modes are unaffected by the gauge field. We further assume equal amplitude scale-free primordial spectra of left- and right handed gravitational waves. These initial conditions, and the assumption that the initial tensor fluctuations of the gauge field vanish deep in the radiation era, allows the YM fluid to behave like radiation at early times. The effects of the gauge field on the subsequent evolution of the gravitational waves are illustrated in Figs \ref{fig:fig1a} and \ref{fig:fig1b}.

The evolution of the gravitational wave amplitude is shown for a variety of cases in Figs.~\ref{fig:fig1a} and \ref{fig:fig1b}. We begin by examining the behavior in the case $g_{\rm YM}=0$, corresponding to color electrodynamics. The background solution has $\phi'$ constant, so the $h_{A}$ evolution equation (where the subscript ``A" is for ambidextrous, since there is no parity violation in this case) has a tachyonic mass that is responsible for the growth of long wavelength modes. As we have shown analytically, $h_A$ doubles for modes outside the horizon, relative to the standard case. As seen in Fig.~\ref{fig:fig1a}, the amplitude of $(a h_{A})^2$ (blue) is $2^2$ times the standard case $(a h)^2$ (green) going in to the first oscillation. We also notice that after modes enter the horizon, there is a slow exchange of amplitude between $h$ and $t$.

%%%%%%%%%%%%%%%%%%%%%%%%%%%%%%%%%%%%%%%%%%%%%%%%%%%%%%%%%%
\subsubsection{Short wavelengths}
 
To investigate further, we make a Wentzel-Kramers-Brillouin (WKB) analysis \cite{Sakurai:2011zz} to obtain the gravitational wave behavior up to high wave numbers -- a configuration space region that is unaccessible numerically due to high computation cost. The behavior of the gravitational wave amplitude as a function of time for different values of the wave number may be obtained by solving Eqn.~(\ref{eqn:colorh}) numerically. However, once a mode enters the horizon, with $k \tau \gg 1$, then the speed of computation slows. In this case, we may adopt the WKB approximation developed here in order to solve for the more slowly evolving envelope of the wave amplitude. We write 
\begin{equation}
h_A = \frac{{\eta}(\tau)}{a} \sin(k \tau + \theta_{\eta}) ,\quad
t_A = \frac{{\nu}(\tau)}{a} \sin(k \tau + \theta_{\nu})
\end{equation}
and define $\Delta\theta  \equiv \theta_\eta - \theta_\nu$. The functions we have to determine are $\eta$, $\nu$ and $\Delta\theta$. In the limit $k \tau \gg 1$ the equations of motion become
\begin{equation}
\begin{aligned}
\eta' &= -k_0\frac{a_0}{a} \nu \cos\Delta\theta, \quad
\nu' = k_0\frac{a_0}{a} \eta \cos\Delta\theta \\
\Delta\theta' &= k_0\frac{a_0}{a} \left( \frac{\nu}{\eta} - \frac{\eta}{\nu}\right) \sin\Delta\theta.
\end{aligned}
\end{equation} 
Hence, our procedure is to solve the full set of equations until $k \tau \gg 1$, at which point we set initial conditions for the envelopes $\eta,\,\nu$ and the phase separation $\Delta\theta$. This set of equations is more easily solved and provides an excellent fit to the full numerical solution. First of all, we notice that the sum $\eta + \nu$ is a constant, indicative of a conserved quantity in the coupled system of gravitational waves and gauge field tensor fluctuations. This visualizes the property of Fig.~\ref{fig:fig1a} nicely: changes in the envelope of $h_A$ are compensated by the envelope of $t_A$. We plot the two solutions together in order to clearly show this effect (see Fig.~\ref{fig:fig1a}). Second, we notice that each time derivative term contains a factor $k_0 a_0/a$, suggesting that the natural time parameter for the modulation of the envelope is $k_0 d\tau a_0/a$. Our solutions show agreement with the numerical results, with the amplitude of $h_A$ and $t_A$ changing with a period defined by $k_0 \int d\tau a_0/a = 2 \pi n$ for $n$ an integer.

The high-frequency behavior of the gravitational waves in the case $g_\ym \neq 0$ is also amenable to a WKB analysis. Upon making the same definitions for $\eta,\,\nu$, the equations of motion become
\begin{equation}
\begin{aligned}
	\eta_{R}' &= -k_0\frac{a_0}{a} \nu_{R} (d\psi \cos\Delta\theta_{R} - \psi^2 \sin\Delta\theta_{R}) \\
	\nu_{R}' &= k_0\frac{a_0}{a} \eta_{R} (d\psi \cos\Delta\theta_{R} - \psi^2 \sin\Delta\theta_{R}) \\
	\Delta\theta_{R}' &= k_g \psi + k_0\frac{a_0}{a}\left( \frac{\nu_{R}}{\eta_{R}}-\frac{\eta_{R}}{\nu_{R}}\right) \\
	&\quad \times (d\psi \sin\Delta\theta_{R} + \psi^2 \cos\Delta\theta_{R}) \\
	\eta_{L}' &= -k_0\frac{a_0}{a} \nu_{L} (d\psi \cos\Delta\theta_{L} + \psi^2 \sin\Delta\theta_{L}) \\
	\nu_{L}' &= k_0\frac{a_0}{a} \eta_{L} (d\psi \cos\Delta\theta_{L} + \psi^2 \sin\Delta\theta_{L}) \\
	\Delta\theta_{L}' &= -k_g \psi + k_0\frac{a_0}{a}\left( \frac{\nu_{L}}{\eta_{L}}-\frac{\eta_{L}}{\nu_{L}}\right) \\
	& \quad \times (d\psi \sin\Delta\theta_{L} - \psi^2 \cos\Delta\theta_{L})
\end{aligned}
\end{equation}
Our procedure is to solve the full set of equations, for both $R-$ and $L-$handed gravitational waves, until $k \tau \gg 1$. Then we use our numerical solution to set initial conditions for $\eta,\, \nu,\, \Delta\theta$ and then evolve the envelopes forward to the present day. The results are qualitatively similar to the ambidextrous case, except that for certain ranges in the value of the wave number $k$ there is a large difference in the amplitudes for $R-$ and $L-$handed gravitational waves. Looking at the figure showing the evolution of the wave amplitude as a function of time (Fig.~\ref{fig:fig1b}), we again see a slow mixing of power between the gravitational wave and the tensor fluctuation of the gauge field. There is a slight boost to the $R-$handed gravitational wave at horizon entry relative to the standard case, and a small suppression of the $L-$handed wave. The WKB approximation is shown to do an excellent job of tracing the shape of the envelope in the fast oscillating regime.

%%%%%%%%%%%%%%%%%%%%%%%%%%%%%%%%%%%%%%%%%%%%%%%%%%%%%%%%%%
\subsubsection{Gravitational wave spectral density}
 
\begin{figure}[h]
\includegraphics[width=0.99\linewidth]{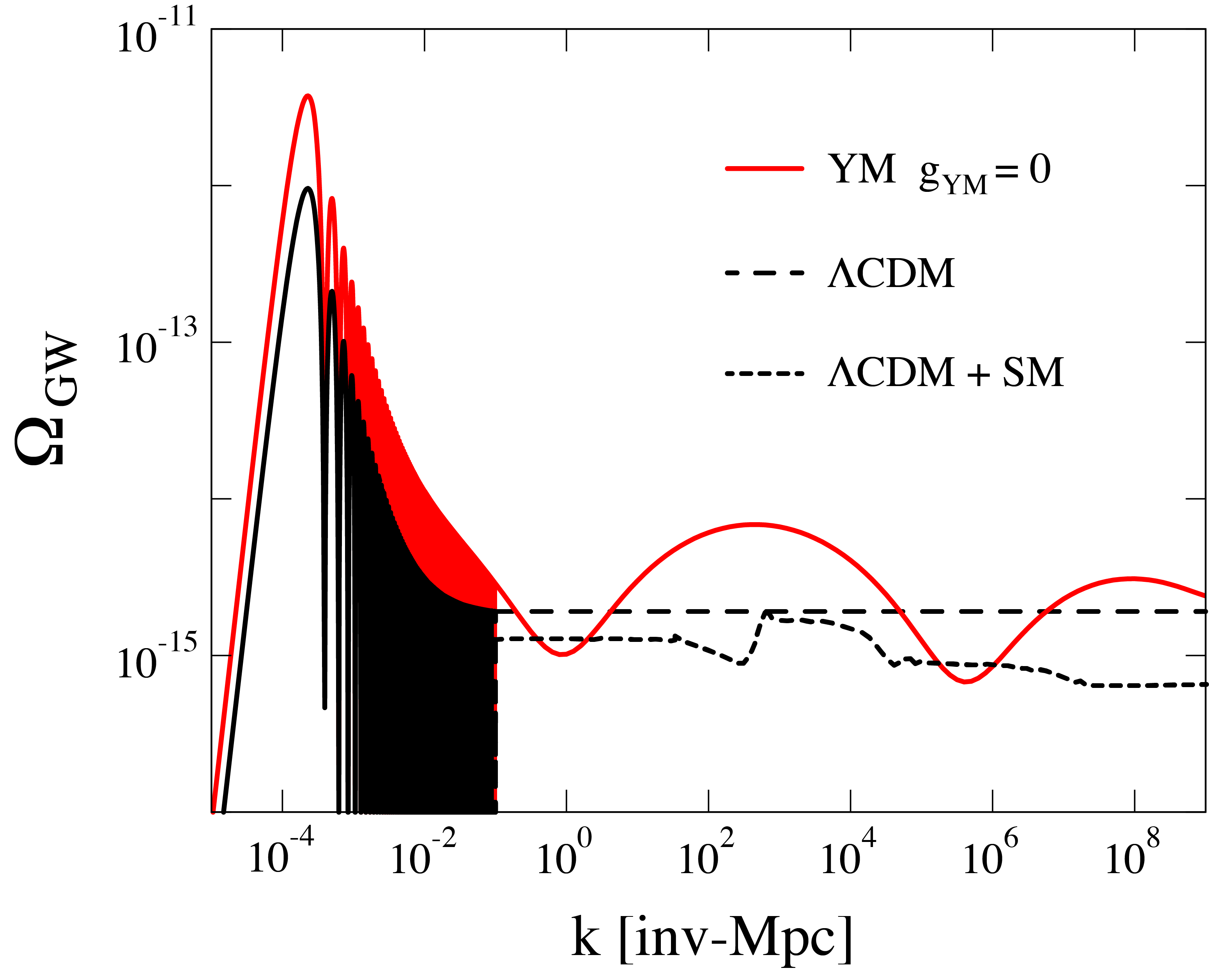}
\caption{The gravitational wave energy density spectrum is shown as a function of the comoving wave number. An ambidextrous, scale free spectrum at an inflationary scale $H_I = 10^{-5}M_P$ is assumed. The present-day spectrum in the case with $g_{\rm YM}=0$, $R_{\rm YM}=0.03$ displays large oscillatory features due to the coupling between the gravitational waves and the gauge field. For comparison, the standard case without the gauge field is shown, as well as the effect of Standard Model particle free streaming and freeze-outs (reproduced from Ref.~\cite{Watanabe:2006qe}).}
\label{fig:fig2a}
\end{figure}

\begin{figure}[h]
\includegraphics[width=0.99\linewidth]{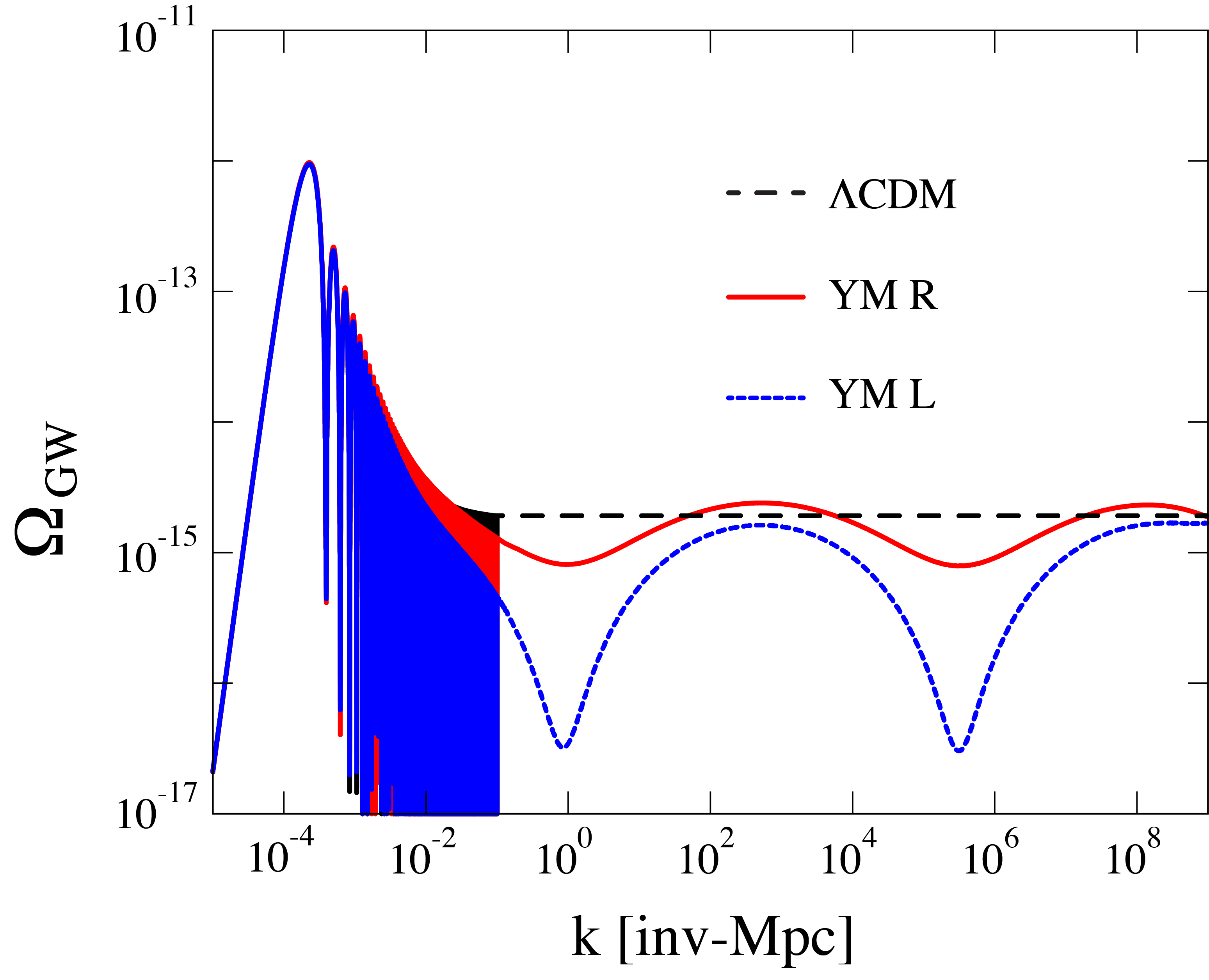}
\caption{As in Fig.~\ref{fig:fig2a}, the gravitational wave energy density spectrum is shown as a function of the comoving wave number, but with $g_{\rm YM}=10^{-60}$, $R_{\rm YM}=0.03$. The long wavelength modes are unaffected given our choice of initial conditions in the background field. However, there is a striking difference in the spectra of $L$- and $R$-handed gravitational waves.}
\label{fig:fig2b}
\end{figure}

The gravitational wave spectral density is the gravitational wave energy density per log frequency interval, in units of the critical density. Following notation in \cite{Watanabe:2006qe}, the spectral density for inflationary gravitational waves is
\begin{equation}
\Omega_{GW} = \sum_{s=L,R}\frac{\Delta_{s}^2}{12 a^2 H^2} T_{s}'(\tau,k)^2
\end{equation}
where the inflationary initial conditions for the amplitude are $\Delta^2_{L,R} = 8 (H_I/M_P)^2 / \pi$ and $T$ is the transfer function defined such that $T(\tau, k) = h(\tau,k) / h(\tau_i,k)$. The gravitational wave spectral density $\Omega_{\rm GW}$ is shown in Figs.~\ref{fig:fig2a} and \ref{fig:fig2b}, where the amplification and periodic modulation are clearly seen. The WKB solution predicts a peak or dip in the spectral density every 5 orders of magnitude in $k$ for $R_{\rm YM}=0.03$; the origin of this oscillation in the spectral density is the same as the oscillation of the amplitude of the envelope of the wave amplitude between the metric and gauge field tensor perturbations.  
 
The difference in the evolution for left- and right-circularly polarized waves is primarily due to the effective mass term for the gauge field tensor perturbations, $-2 k g_{\rm YM}\phi$, which is tachyonic for right handed modes. This behavior leads to an interesting effect: The growth (suppression) of $t_{R}$ ($t_{L}$) is transferred to $h_{R}$ ($h_{L}$) as the mode enters the horizon (see Fig.~\ref{fig:fig1b}). Once the relative amplitude is locked in at horizon entry and the fields begin to oscillate rapidly, the slow exchange of amplitude between $h_{R/L}$ and $t_{R/L}$ again comes into play. The WKB analysis for subhorizon modes shows that $h_{R/L}^2 + t_{R/L}^2 \propto 1/a^2$ and the exchange is oscillatory with similar phase if the background field is not yet oscillatory. 
 
The WKB solution allows us to easily calculate the gravitational wave spectral density out to high frequencies. Looking at Fig.~\ref{fig:fig2a}, and comparing with the standard case  obtained from \cite{Watanabe:2006qe} we notice several significant features. First, the long wavelength amplitude is higher by a factor of 4 than in the standard case, as expected. Second, there are slow, secular oscillations in the amplitude, as predicted based on the envelope oscillation phase $a_0 k_0 \int d\tau /a$. The decay of the oscillation amplitude at high-frequency is due to the choice of starting time for our integrations. Those wave modes start outside the horizon, but did not reach the amplitude-doubling fixed point before horizon entry. We started these integrations at a redshift $z \sim 10^{16}$ for reasons of numerical accuracy. If we had started at a more physically realistic $z \sim 10^{27}$, then the decay of oscillations in $\Omega_{GW}$ would be shifted to even higher frequencies. Third, and finally, for the values of the parameters $g_\ym$ and $R_\ym$ used in this figure, the oscillations in the amplitude of $\Omega_{GW}$ due to the coupling with the gauge field are much larger than the imprint of the thermal history of the relativistic fluid as calculated in \cite{Watanabe:2006qe}.

The gravitational wave spectral density for the case of the fully nonlinear Yang-Mills gauge field is shown in Fig.~\ref{fig:fig2b}. Here we see the striking asymmetry in the left- and right handed gravitational wave spectrum near $k \sim 10^0,\, 10^5$~inv-Mpc. Although our numerical calculation extends only to $k = 10^9$~inv-Mpc, the asymmetric modulation of the two spectra should continue to higher wave numbers. Should the direct detection of a stochastic gravitational wave background ever  threaten to become a reality, one might consider attempting to distinguish between the left- and right handed spectrum. Finally, we reiterate that the details of our spectrum depend on choices of initial conditions for the tensor modes $t_{L,R}$ (which we have set to zero) and the background field $\phi$ (in terms of the split between electric and magnetic energy).

%%%%%%%%%%%%%%%%%%%%%%%%%%%%%%%%%%%%%%%%%%%%%%%%%%%%%%%%%%
\subsection{Vector modes}
\label{subsec:vector}

Here we examine the behavior of the vector modes, based on Eqns.~(\ref{eqn:vecpert}). We solve for the constraint modes $Y_I$ as described earlier, leaving second-order equations for $M_I$ that are sourced by $B_I$.

For our first attempt to study this complicated system of equations, we set $B_I=0$. The $M_I$ differential equations are coupled, so we make the transformation
\be
   M_1 = 2g_\ym^3 k^2 a^2 f_1 + i g_\ym^2 k a^2\left( k^2 + 2g_\ym ^2 \phi^2\right)\frac{f_2}{\phi}
\ee
to unmix the second-order terms, where $f_I$ are the new vector perturbation variables and again, the mapping $\{1,2,i\} \to \{2,1,-i\}$ yields the second equation. The new equations are still rather complicated, so we consider the limiting cases of very high and very low frequencies. In the case $k \gg g_\ym \phi$, the equation of motion becomes
\be
f_I'' + 6 \frac{a'}{a} f_I' + k^2 f_I=0
\ee
which describes a damped harmonic oscillator with the damping depending on the cosmology. In the case of power-law expansion, with $a \propto \tau^{2/(1+3 w)}$ where $w$ is the background equation of state, then $f_I \propto j_n(k\tau)/\tau^n$ and $n=(5-3w)/(1+3w)$. To estimate the consequences of this solution, one would typically compute the density contrast $\delta \rho / \rho$ or a similar quantity to see if it grows or decays. However, in this specific gauge choice of Eqn.~(\ref{eqn:perts}) the energy density and pressure perturbations only appear at second-order. The off-diagonal terms in the YM stress energy tensor are first-order in perturbations though, so we compare the off-diagonal terms to the unperturbed energy density to determine if the perturbations grow. For example, the dominant contributor to the momentum density  is $\delta T_{tx} \sim -i g_\ym k M_2 \phi \phi'/a$ and $M_1 \sim i g_\ym^2 a^2 k^3 f_2/\phi$. Using our analytic solution, and $T_{tt} = a^2 \rho_\phi$ then
\be
\frac{\delta T_{tx}}{T_{tt}} \simeq \frac{2 g_\ym^3 k^4 \phi'}{3(\phi'^2 + g_\ym^2 \phi^4)}a^3 \tau^{-n} j_n(k \tau) e^{ikz}.
\ee
Using the asymptotic behavior of the spherical Bessel function, $j_n(x) \to \frac{1}{x}\cos(x-(n+1)\pi/2)$ as $x\to \infty$, then we see that the momentum density conveyed by the vector perturbation oscillates with constant profile, $\delta T_{tx} / T_{tt} \propto e^{i k \tau}$. Similar results are obtained for other nonzero off-diagonal components of the stress energy tensor. Hence, there is no gravitational instability for modes with $k \gg g_\ym \phi$. Based on typical numbers used in our study, this translates to $g_\ym \phi \sim 10^{-3} H_0$. For practical purposes, this applies to all modes of interest.

Next we include the vector metric perturbations, by solving the algebraic constraints for $B_I$ and $Y_I$, leaving second-order equations for $M_I$. The same transformation is made to unmix the the second-order terms in favor of the variables $f_I$. In this case, the high-frequency limit for the evolution of $f_I$ is the same as before. Hence, our analysis from above still holds true, that the high-frequency vector perturbations oscillate with constant profile.

To study the very low frequency vector perturbations, we start again with the case $B_I=0$ and proceed to unmix the equations of motion. Taking the leading terms in the limit $k \ll g_\ym \phi$, then the equation of motion for $f_I$ reduces to $(a^3 \phi f_I)'' = 0$. Since the dominant contributor to the momentum density perturbation in this limit is $\delta T_{tx} \sim i g_\ym^4 k^3 (a^3 \phi^2 f_I)' /a^2$, then in units of the energy density we find
\be
\frac{\delta T_{tx}}{T_{tt}}  \simeq \frac{2 i g_\ym^4 k^3 e^{ikz}}{3(\phi'^2 + g_\ym^2 \phi^4)} \left(c_1 \phi' + c_2 (\tau \phi' + \phi)\right) \, ,
\ee
where $c_{1,2}$ are integration constants. This growing term appears to indicate an instability due to the linear growth in $\tau$; however, it is simply the secular perturbation discussed in the prologue to this section. That is, by adjusting the initial values of $\phi$ and $\phi'$ then the constants $c_{1,2}$ can be made to vanish.  When we include the vector metric perturbations, once again the same equations of motion and momentum density perturbation are obtained in the small-$k$ limit. There is no instability.

Because there is no external source of vector perturbations, and because the YM vector does not amplify any perturbations, we have chosen to omit computing the contributions of vector perturbations to the CMB anisotropy in Sec.~\ref{sec:cmb}.

%%%%%%%%%%%%%%%%%%%%%%%%%%%%%%%%%%%%%%%%%%%%%%%%%%%%%%%%%%
\subsection{Scalar perturbations} 
\label{sec:scalar}

The scalar perturbations are stable, as we now illustrate. The equations of motion for the scalar degrees of freedom, $\delta\phi$ and $\delta m$, can be simplified by making the change of variables
\begin{equation}
\delta\phi = \Delta_1,\, \delta m = \Delta_1 - \Delta_2 \sqrt{k^2 + 2 g_\ym^2 \phi^2} /(g_\ym \phi).
\end{equation}
In this case, the second-order terms in the equations of motion unmix, so that
\be
\Delta_i'' + {\cal M}_{ij} \Delta_j = \Sigma_i
\ee
where
\begin{align*}
& {\cal M}_{11} =  k^2+ 4 g_\ym^2 \phi^2 \\
& {\cal M}_{12} = {\cal M}_{21} =  2 g_\ym \phi \sqrt{k^2 + 2 g_\ym^2 \phi^2} \\
& {\cal M}_{22} =  k^2 + 2 g_\ym^2\phi^2 + \frac{2 k^2 g_\ym^2\phi^2}{k^2 + 2 g_\ym^2\phi^2} + \frac{6 k^2 g_\ym^2\phi'^2}{(k^2 + 2 g_\ym^2 \phi^2)^2}\\
& \Sigma_1 = 4 g_\ym^2 \phi^3 \Phi_G - \phi' \Phi_G' - k^2 \phi' b \\
& \Sigma_2 = -  \frac{(3 k^2 + 2 g_\ym^2\phi^2)g_\ym\phi \phi'}{(k^2 + 2g_\ym^2\phi^2)^{3/2}}k^2 b 
- \frac{g_\ym\phi^2}{k^2 + 2 g_\ym^2\phi^2}k^2 b' \\
&+ \left(\frac{(k^2+4 g_\ym^2\phi^2)g_\ym \phi^2}{\sqrt{k^2 + 2 g_\ym^2\phi^2}} - \frac{2 k^2 g_\ym \phi'^2}{(k^2+2 g_\ym^2\phi^2)^{3/2}} \right)\Phi_G\\
&-  \frac{g_\ym\phi'\phi}{\sqrt{k^2 + 2 g_\ym^2\phi^2}} \Phi_G' .
\end{align*}
These are just two coupled, driven harmonic oscillators. The eigenvalues of the matrix ${\cal M}$ are positive definite, so the oscillator is stable. At long wavelengths, however, we can see the imprint of the secular instability. In the $k \to 0$ limit, the homogeneous equations become $\delta m''=0$ and $\delta u'' + 6 g_\ym^2 \phi^2 \delta u=0$ where $\delta u = 3\delta\phi - \delta m$. Of course, we recognize the $\delta u$ equation as Eqn.~(\ref{eqn:linsec}). When we insert the growing solution (\ref{eqn:linsecsol}) into, say, the energy density in Eqn.~(\ref{eqn:fluidvars}) then we find $\delta\rho/\rho \propto \delta u'(\tau_i)$ is a constant and ought to be absorbed into the background. Similarly, the shear appears to grow linearly
\be
\sigma = 2\frac{g_\ym^2\phi^3(\delta m(\tau_i) + \delta m'(\tau_i) \tau) - \phi' \delta m'(\tau_i)}{3(\phi'^2 + g_\ym^2 \phi^4)}
\ee
but again this is an artifact of our linearization. The system is stable as shown by our numerical calculations in the next section.

%%%%%%%%%%%%%%%%%%%%%%%%%%%%%%%%%%%%%%%%%%%%%%%%%%%%%%%%%%
\section{CMB} 
\label{sec:cmb}

Ultimately, detecting the imprint of this chiral YM fluid requires an analysis of the CMB sky. In this section we compute the power spectra and describe ways to detect deviations from vanilla $\Lambda$CDM. To evaluate the impact of this scenario on the CMB, we have implemented the scalar and tensor perturbations of the gauge field into CAMB \cite{Lewis:1999bs}. We implement the standard adiabatic perturbation for scalar and tensor perturbations but set $t(\tau_i) = 0$ lacking an early Universe theory for this toy model. This leaves us with solutions that live close to vanilla \lcdm so these new effects are not artificially enhanced. The gauge field has the biggest impact on tensor correlations. The scalar sector also receives corrections due to the gauge field, in the form of an anisotropic scalar shear, but the impact on the scalar CMB spectrum is small. We ignore the vector perturbations which we showed to decay in Sec.~\ref{sec:vec}. We use the parameter $R_{\rm YM}$ describing the ratio between the YM fluid density and the total relativistic energy density. We assume the fraction of critical density in the relativistic fluid is fixed by slightly adjusting the sum of the neutrino masses upon introducing the gauge field. This avoids simply comparing the YM fluid in terms of an increased effective neutrino number $N_{\rm eff}$ as we are trading the sum of total neutrino mass for $\Delta N_{\rm eff}$. We otherwise assume standard $\Lambda$CDM parameters. Details of our implementation can be found in Appendix \ref{appendix}.

\begin{figure}[htbp]
	\includegraphics[width=1\linewidth]{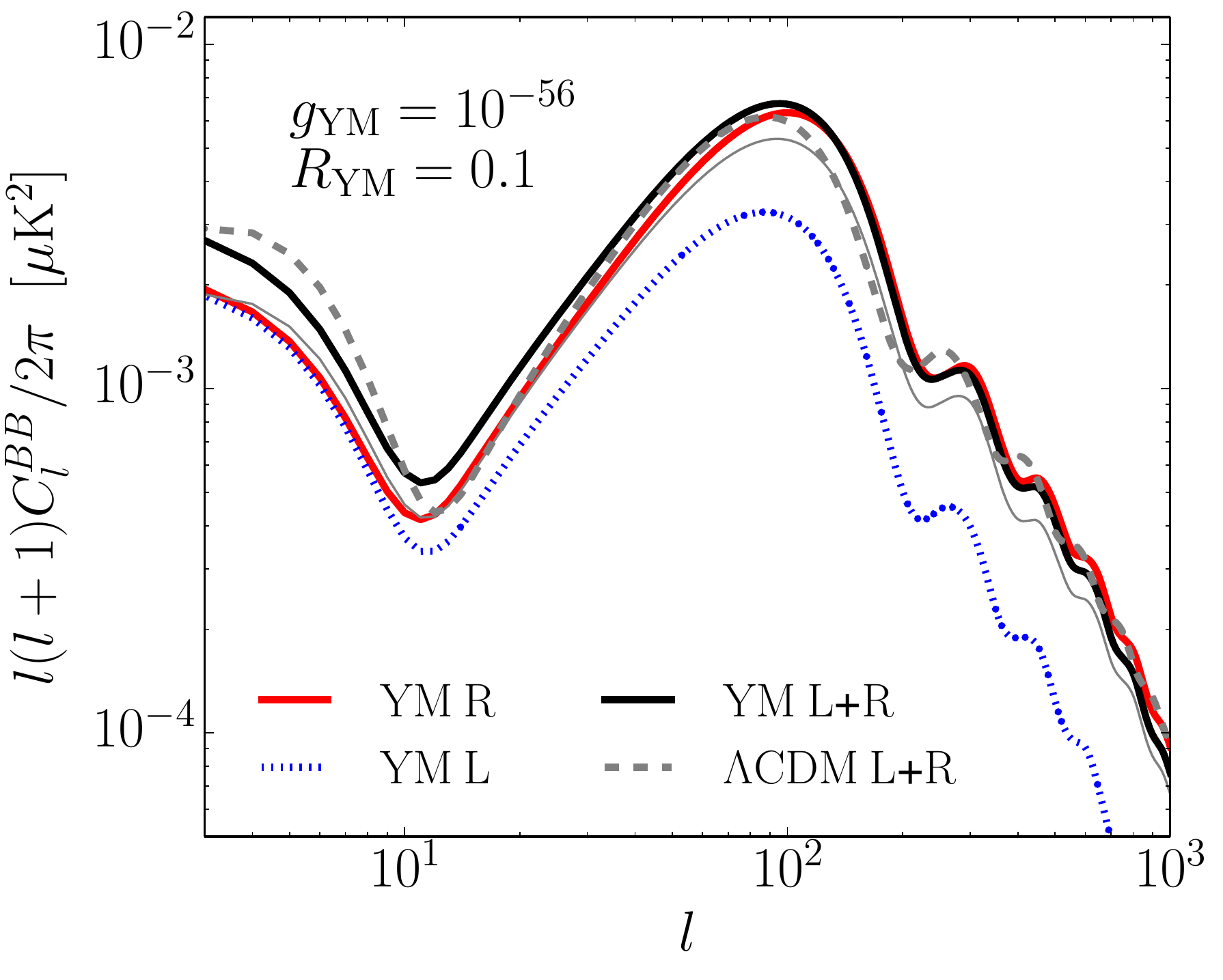}
	\caption{CMB $BB$ polarization autocorrelation spectra. The pure left- and right handed contributions deviate strongly from $\Lambda$CDM (dashed) while their sum is closer to it. Solid lines include a gauge field with $g_{\rm YM} = 10^{-56}$, $R_{\rm YM} = 0.1$ and tensor-to-scalar ratio of $r = 0.1$. The thin line shows the $BB$ spectrum for the case $g_{\rm YM} = 4 \times10^{-56}$.}
	\label{fig:BB}
\end{figure}

\begin{figure}[htbp]
	\includegraphics[width=1\linewidth]{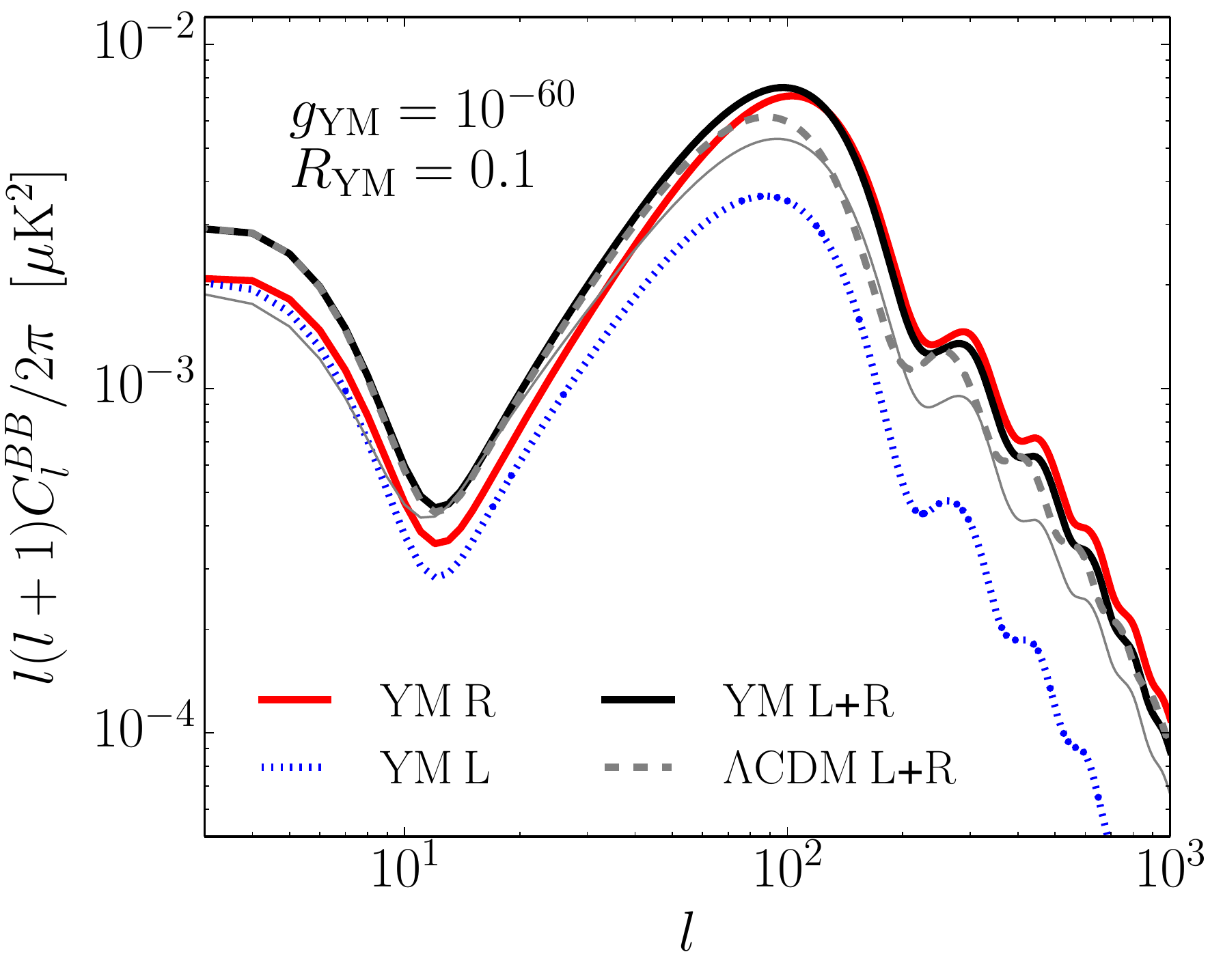}
	\caption{Graph analogous to Fig.~\ref{fig:BB}: the gauge field now has $g_{\rm YM} = 10^{-60}$, $R_{\rm YM} = 0.1$ and tensor-to-scalar ratio of $r = 0.1$. Again, the thin line shows the $BB$ spectrum for the case $g_{\rm YM} = 4 \times10^{-56}$.}
	\label{fig:BBneg60}
\end{figure}

\begin{figure}[htbp]
	\includegraphics[width=1\linewidth]{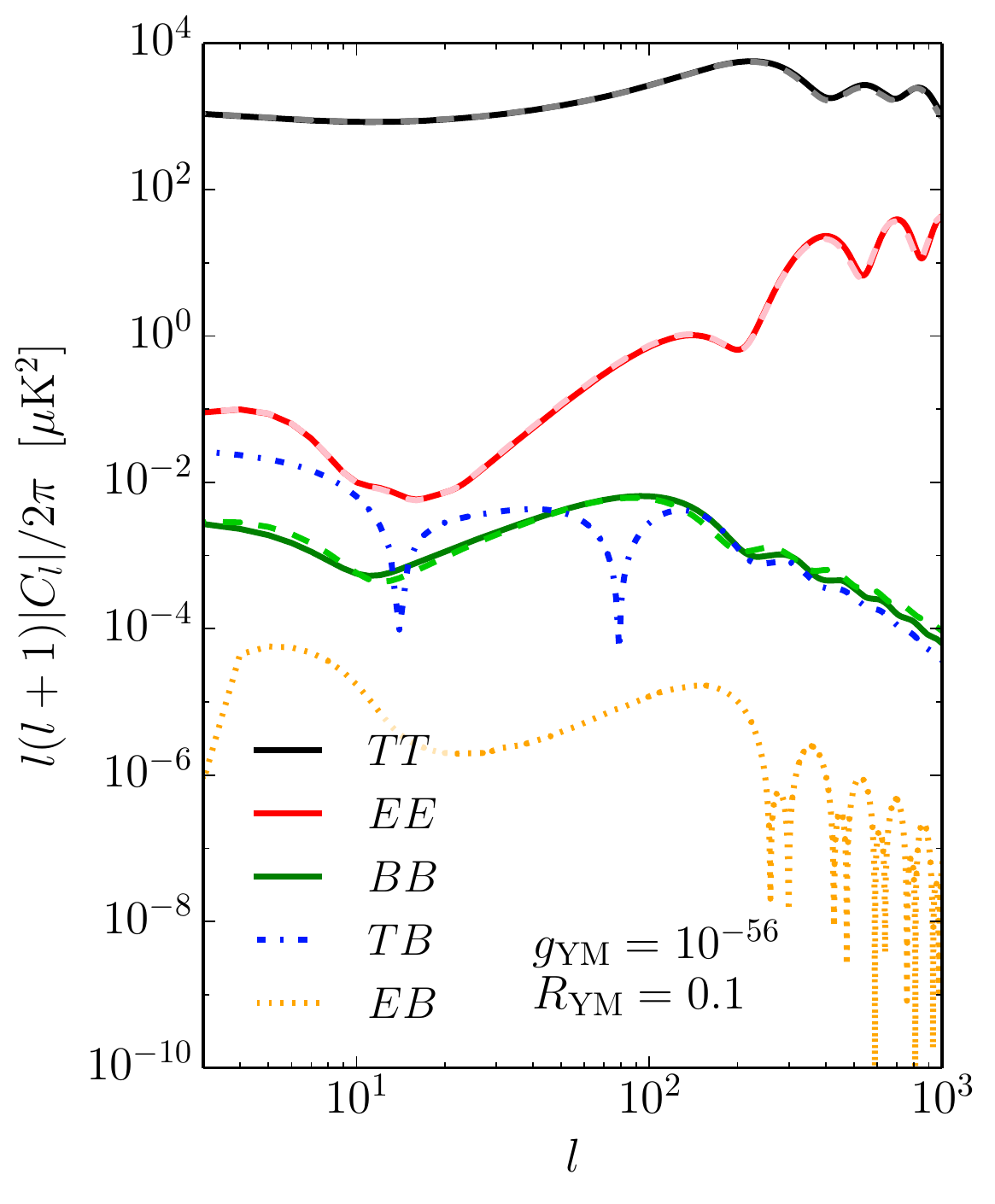}
	\caption{CMB temperature and polarization auto- and cross-correlation power spectra. The parity violation allows for $TB$ and $EB$ cross-correlations. The dashed lines represent the standard $\Lambda$CDM cosmology, solid lines include a gauge field with $g_{\rm YM} = 10^{-56}$, $R_{\rm YM} = 0.1$, $\theta = \pi/4$ and tensor-to-scalar ratio of $r = 0.1$. The $TB$ and $EB$ cross-correlations only appear when the gauge field is present.}
	\label{fig:tens}
\end{figure}

The CMB polarization can be decomposed into gradient $E$-modes and curl $B$-modes. In the tensor sector the gauge field introduces two main effects. First, the left- and right handed contributions to the $BB$ spectrum now differ, as shown in Fig.~\ref{fig:BB}. Hence, the temperature and polarization anisotropy due to gravitational waves on roughly degree scales is dominated by a superposition of right-circularly polarized gravitational waves, which imprint left-helical patterns on the sky.
It is curious to see that the individual contributions deviate strongly from vanilla $\Lambda$CDM but conspire in a way that puts the combination of both close to the expected standard cosmology result. This holds true for a wide parameter range, and these effects can be seen in Fig.~\ref{fig:BBneg60}, the corresponding graph with $g_\ym = 10^{-60}$. The general features in this parametrization are similar, although the YM modified spectrum follows the vanilla \lcdm one more closely, especially at large scales. Second, because temperature $T$ and gradient polarization $E$ are both parity even but curl polarization is parity odd, the correlation spectra between these types vanish in the vanilla $\Lambda$CDM scenario. However, the parity violation introduced by the YM fluid allows for correlations between these types, $TB$ and $EB$ \cite{Lue:1998mq}. Detecting these exotic cross-correlations is a smoking gun for chiral effects in the Universe. Typical predictions of our model are plotted in Fig.~\ref{fig:tens}. In that plot we include all types of correlations to make a comparison between the signal strength of the different spectra easier. $\Lambda$CDM spectra are plotted with dashed lines. 

\begin{figure}[htbp]
	\includegraphics[width=1\linewidth]{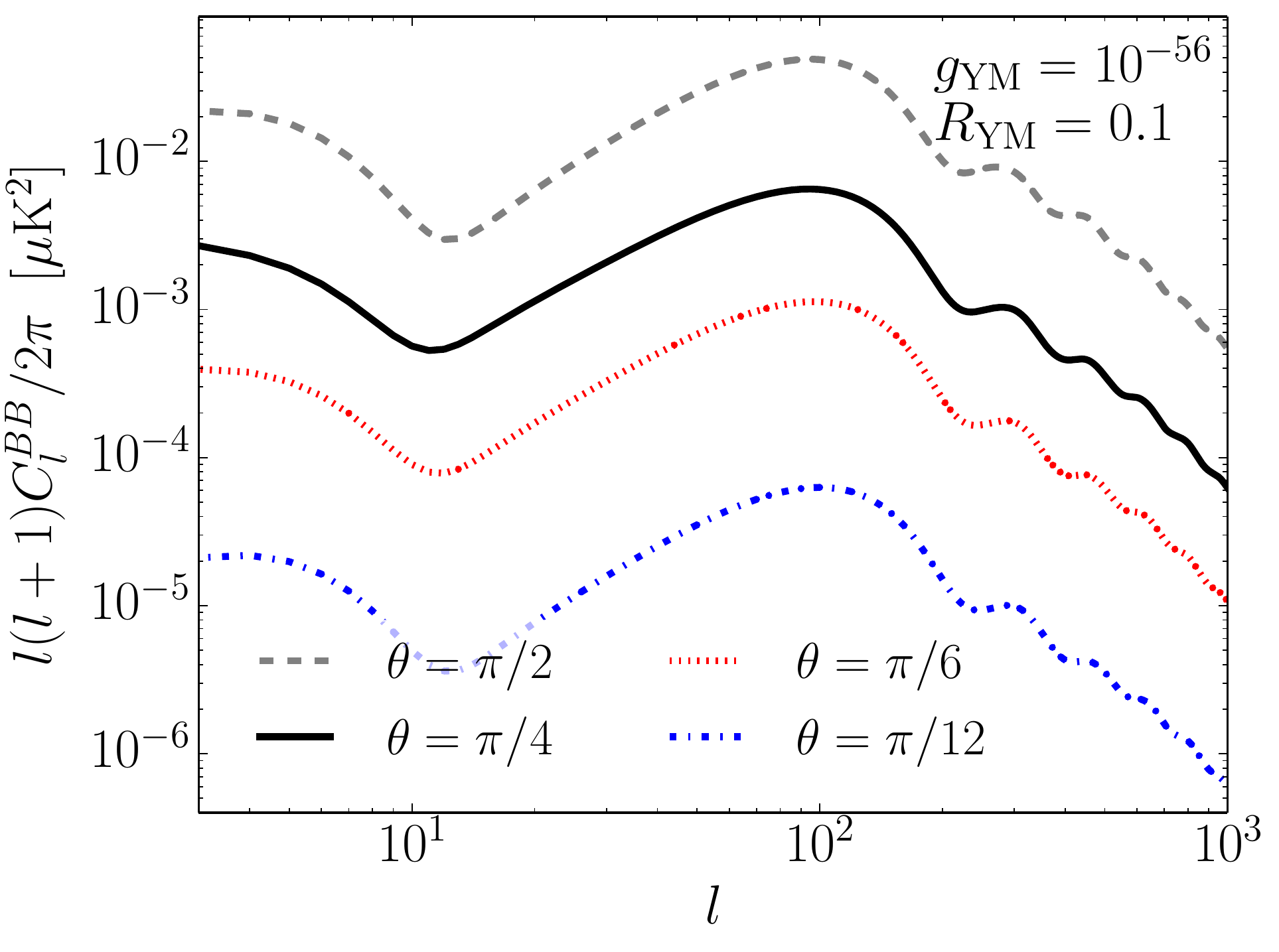}
	\caption{Curl polarization power spectra. The cosmological parameters as well as $\{g_\ym, R_\ym\} = \{10^{-56}, 0.1\}$ are kept constant. The angle $\theta$ is being varied which changes the contribution of initial electric and magnetic YM energy. The $\theta = \pi/4$ curve is closest to $\Lambda$CDM.}
	\label{fig:BBvarytheta}
\end{figure}

So far this discussion was centered around choosing initial conditions that resemble vanilla $\Lambda$CDM. In particular, the angle $\theta$ in Eqns.~(\ref{eqn:initial}, \ref{eqn:transfer}) was chosen to be equal to $\pi/4$. Deviating from this has two effects: The background evolution of $\phi$ changes drastically and the effective mass term in the evolution equation for $h$ no longer vanishes as described in Sec.~\ref{sec:longwave}.
This behavior is shown in Fig.~\ref{fig:BBvarytheta}. The solid black line describes the parametrization that has been employed in the rest of this paper, using $\theta = \pi/4$. Going to larger angles $\theta \to \pi/2$ lifts the entire $BB$ power spectrum almost by an entire order of magnitude. Half of the logarithmic increase comes from the new initial conditions, as evidenced by Eqns.~(\ref{eqn:transfer}). We are using this equation to set up initial conditions for CAMB at $a = 10^{-8}$. The other half is caused by the change in the background field behavior. Propagation of the tensor perturbations through this altered background field yields this strong effect on the spectra. \emph{De}creasing the angle lowers the spectrum indefinitely. This analysis shows the extent to which $\theta$ changes the behavior of the YM fluid and how sticking to $\theta = \pi/4$ gives results closest to vanilla $\Lambda$CDM. However, to explore the entire parameter range, we need to take variations in the distribution of initial electric and magnetic energy contribution into consideration.

There are many challenges to detecting the parity-violating cross-correlations, not to mention the primordial $B$-mode signal. Galactic foregrounds, magnetic fields, weak lensing, and other systematic effects can all produce a false positive; fortunately there is no fundamental barrier that would prevent a detection that can distinguish a primordial signal. (See Ref.~\cite{Kaufman:2014rpa} for a recent summary.) 

But there are other, competing, phenomena that could produce a parity-violating signal. First of all, cosmological birefringence (CB) can lead to $TB$ and $EB$ power spectra by rotating $E$ into $B$ through a novel coupling between electromagnetism and a cosmic pseudoscalar such as quintessence \cite{Carroll:1998zi}.  A second possibility, broadly characterized as chiral gravity, posits a modification of gravity whereby an asymmetry between left- and right-circularly polarized waves is imprinted on the primordial spectrum. The third possibility, as we have shown, is essentially cosmic circular dichroism, whereby the asymmetry develops with time from an initially symmetric primordial spectrum.

Would an actual detection directly point to chiral symmetry breaking on cosmological scales? Do the birefringence effects mask true chiral physics? In Ref.~\cite{Gluscevic:2010vv} it was shown that the $TB$ and $EB$ spectra can be used to distinguish these CB effects from chiral physics. As CB rotates the $E$ into a $B$ contribution the measured $B$ spectrum would resemble the $E$ one which makes this separation into CB and chirality effects feasible. In turn, putting limits on the amplitude of these spectra will put constraints on chiral physics in general and our model in particular. 

In what follows we compute the constraints that current and future CMB experiments would put on the parameters of our model under the assumption that $TB$ and $EB$ cross-correlations are measured.

%%%%%%%%%%%%%%%%%%%%%%%%%%%%%%%%%%%%%%%%%%%%%%%%%%%%%%%%%%
\section{Fisher Forecasts} 
\label{sec:forecasts}
	
The deviations seen in the CMB spectra would clearly have an impact on the interpretation of a precision measurement of $B$ modes \cite{Ade:2014xna}.  
As can be seen from Fig.~\ref{fig:BB}, the gauge field can vary the height of the $BB$ spectrum at the reionization bump near $\ell \lesssim 10$ and at the primary acoustic peak near $\ell \sim 100$ by as much as $\pm 50\%$. However, we have the greatest leverage on new physics by focusing on the exotic cross-correlations. Hence, we forecast the parameter constraints $\sigma_{R_{\rm YM}}$ and $\sigma_{g_{\rm YM}}$ using these spectra. 

The Fisher-matrix technique gives an estimate for the statistical error in a given measurement. Therefore, systematic deviations are solely contained in the choice of a fiducial model along with any theoretical bias effects. The Fisher matrix for this case reads
\begin{equation}
	\mathcal{F}_{ij}=\sum_l \sum_{X,Y}\frac{\partial C_l^X}{\partial p_i}\frac{\partial C_l^Y}{\partial p_j}\left[\Xi^{-1}_l\right]_{XY}
	\label{eqn:fisher}
\end{equation}
where $\vec p = (R_{\rm YM}, g_{\rm YM}) + \vec p_{\rm cosmo}$ and $X, \, Y = \{TB, EB\}$ and $\Xi$ is the $C_l$ covariance matrix. Here, the eight cosmological parameters are $\vec p_{\rm cosmo} = (\omega_b, \omega_c, \omega_\nu, \Omega_K, H_0, w, n_t, r)$. (In CAMB, this corresponds to $\texttt{tensor\_parameterization}=1$.) The Fisher matrix $\mathcal{F}$ is the inverse of the covariance matrix between $R_{\rm YM}$ and $g_{\rm YM}$. The derivatives of the $C_l^X$ are obtained using CAMB. We center the derivatives around the following fiducial model: We choose $g_{\rm YM} = 10^{-56}$, $R_{\rm YM} = 0.1$, and the standard Planck $\Lambda$CDM values for the cosmological parameters \cite{Ade:2013zuv}. Furthermore, we set the angle that distributes initial energy contributions between electric and magnetic parts of the YM fluid $\theta = \pi/4$ (Eqn.~(\ref{eqn:initial})).

The matrix $\left[\Xi^{-1}_l\right]_{XY}$ is the inverse of the $TB$-$EB$ covariance matrix given by 
\bes
	\Xi ^{{X_1}{X_2},{X_3}{X_4}}_{l} = \frac{\tilde C_l^{{X_1}{X_3}}\tilde C_l^{{X_2}{X_4}} + \tilde C_l^{{X_1}{X_4}}\tilde C_l^{{X_2}{X_3}}}{2l+1}
\ees 
where 
\bes
	\tilde C_l^{XX'} \equiv C_l^{XX'} + w_{XX'}^{ - 1}|W_l^b{|^{ - 2}}
\ees 
and $X=\{T, \, E, \, B\}$ \cite{Gluscevic:2010vv}, where $W$ is the window function and $w$ describes the instrumental noise. Note the usage of the superscripts: Here $X$ refers to one type of perturbation only, whereas above it describes the cross-correlations. The properties of the detector are imperative in the determination of the parameter constraints. The instrumental parameters enter the window function in two ways: due to the beam width via 
\bes
	W_l^b \simeq \exp \left(-l^2 \sigma^2_b /2\right)
\ees
and the instrumental noise $w_{XX}^{-1}$, where 
\bes
	w_{TT}^{ - 1} \equiv  4\pi \sigma _T^2/N_{\text{pix}}
\ees
and $w_{EE}^{ - 1} = w_{BB}^{ - 1} \equiv  4\pi \sigma _P^2/N_{\text{pix}}$ with the cross-correlation contributions vanishing as the noise in the polarization is assumed to have no correlation to the noise in the temperature. 

In the window function $\sigma_b \equiv \theta_{\rm FWHM}/\sqrt{8\ln 2}$ where the beam width is measured in radians. Similarly, the number of pixels is $N_{\rm pix} = 4\pi\theta_{\rm FWHM}^{-2}$ and $\sigma_T$ and $\sigma_P$ are the temperature and polarization pixel noise. These are given by $\sigma_T^2 = ({\rm NET})^2 N_{\rm pix}/t_{\rm obs}$ and $\sigma_P = \sqrt{2}\sigma_T$ with NET being the noise-equivalent temperature and $t_{\rm obs}$ being the observation time. 

The parameters for this analysis are taken from Ref.~\cite{Gluscevic:2010vv, Planck:2013cta} and summarized in Table~\ref{tab:obs}.
\begin{table}[htbp]
    \begin{tabular}{l c c c}
    \hline
    Instrument \,\,& \,$\theta _{{\rm FWHM}}$ [arcmin] & \,NET \,$[\mu \text{K}\sqrt{\text{s}}]$\, &  \,$t_{\text{obs}}$ [y] \\ \hline
	 Planck & 7.1 & 45 & 2 \\
	 CV limited & 5 & 0 & 1.2 \\ 
    \hline
    \end{tabular}
\caption{Instrumental parameters for the two experiments considered in this paper.  The parameters are the beamwidth $\theta _{\rm{FWHM}}$, noise-equivalent temperature NET, and observation time $t_{\text{obs}}$.}
\label{tab:obs}
\end{table}

\begin{figure}[htbp]
	\includegraphics[width=1\linewidth]{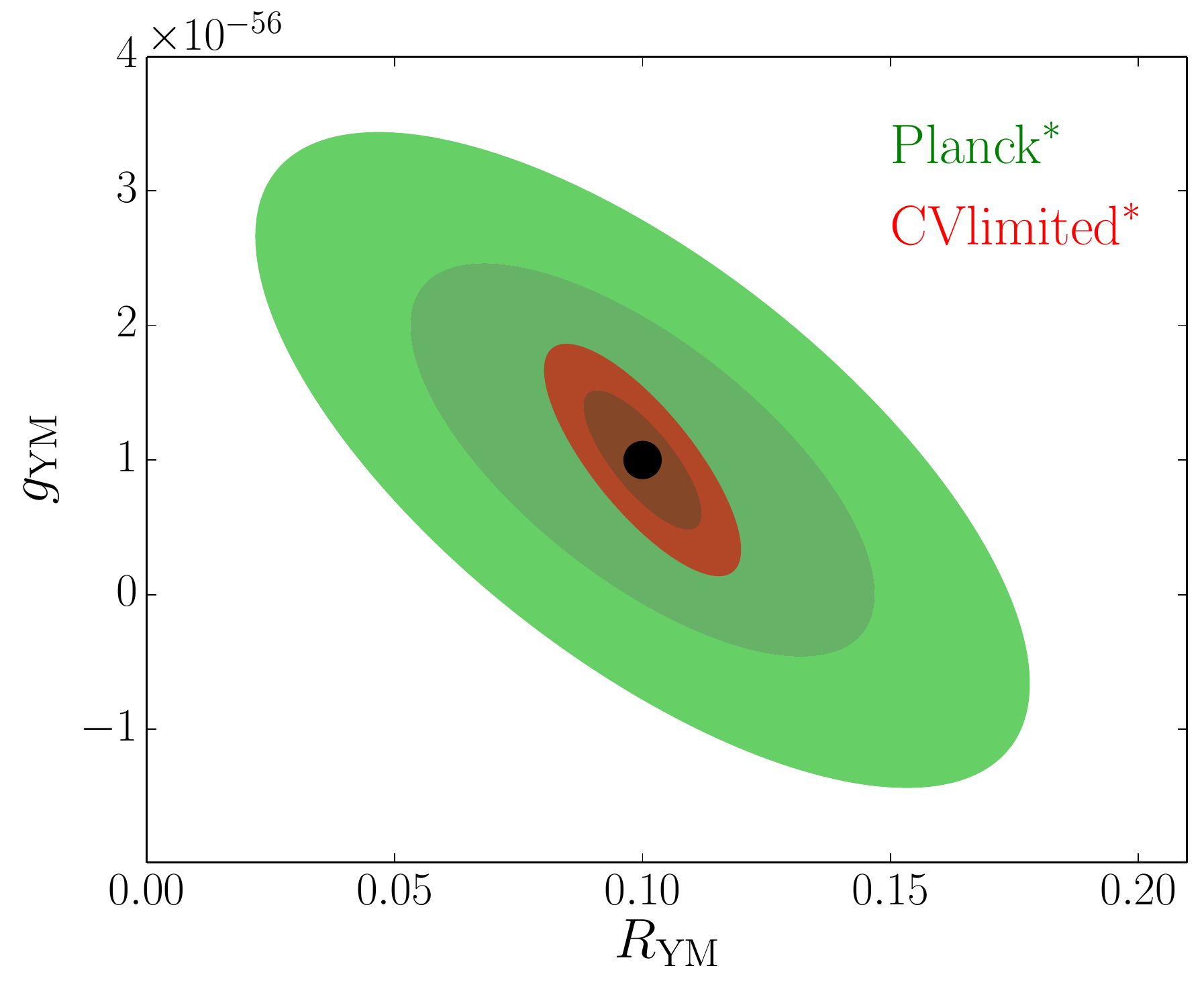}
	\caption{Forecasted 1- and 2-$\sigma$ C.L. contours under the condition that $TB$ and $EB$ cross-correlations are detected by the respective experiments (indicated by the asterisk $*$). The fiducial model is indicated by the black dot which represents $g_{\rm YM} = 10^{-56}$ and $R_{\rm YM} = 0.1$}
	\label{fig:fisher}
\end{figure}

The 1D marginalized confidence limits in a scenario in which the Planck satellite measures $TB$ and $EB$ correlations are $\sigma_{\rm g_{\rm YM}} = 9.5\times 10^{-57}$ and $\sigma_{R_{\rm YM}} = 0.030$. For the cosmic variance (CV) limited experiment these numbers reduce to $\sigma_{\rm g_{\rm YM}} = 3.4\times 10^{-57}$ and $\sigma_{R_{\rm YM}} = 8.1\times 10^{-3}$, which would enable us to make a nonzero detection of $g_{\rm YM}$. The 1- and 2-$\sigma$ contours are plotted in Fig.~\ref{fig:fisher} where we marginalize over the other parameters.
For this fiducial model Planck could make a 2-$\sigma$ detection of $R_{\rm YM}$, but cannot exclude the ambidextrous case since $g_{\rm YM} = 0$ lies within its 1-$\sigma$ contour. The future looks brighter for proposed satellite missions that gets closer to a CV limited experiment (Prism \cite{Andre:2013nfa}, CMBPol \cite{Baumann:2008aq}), which could put constraints on the chiral asymmetry; for the fiducial model, the coupling $g_{\rm YM}$ could be distinguished from zero at better than the 2-$\sigma$ level.

%%%%%%%%%%%%%%%%%%%%%%%%%%%%%%%%%%%%%%%%%%%%%%%%%%%%%%%%%%
\section{Conclusions and Outlook} 
\label{sec:concl} 

We present a simple model of a dark fluid that breaks chiral parity on cosmological scales. We illustrate the impact of this model on the gravitational wave spectrum and the CMB, computing the power spectra along with new $TB$ and $EB$ correlations that emerge in parity-breaking models. For a wide range of parameters the observables lie within current experimental bounds. However, upcoming experiments will be able to put stringent constraints on the $TB$ and $EB$ modes and a detection of one of these correlations could be the sign of a flavor-space locked gauge field. 

The perturbations in all three sectors are bound on all scales. We discover a secular instability in the scalar and vector perturbations of the theory, nonetheless, this is an artifact of a poor choice of initial conditions and does not describe a breakdown of the model. We analyze the effect of changing the YM coupling constant, the total amount of YM fluid and the initial distribution of energy in the electric and magnetic part of the fluid. Varying these initial conditions leads to big changes in the $B$ cross and autocorrelation spectra.

We compute constraints that future experiments put on the parameters of our theory assuming that $TB$ and $EB$ cross-correlation are measured. A detection of this coupling could be used to determine whether the gauge field is part of a dark sector that includes dark energy. We note that the effect of the flavor-space locked SU(2) electric or magnetic fields on particles with charge $g_\ym$ under the same SU(2) group would be negligible. However, if dark energy couples to the rolling gauge field, or if the gauge field is dark energy, as in a gauge-flation scenario, then the rate of cosmic acceleration may be linked to the chiral asymmetry.

%%%%%%%%%%%%%%%%%%%%%%%%%%%%%%%%%%%%%%%%%%%%%%%%%%%%%%%%%%
\acknowledgments 

This work is supported in part by DOE Grant No. DE-SC0010386. J.B. acknowledges support by the Gordon F. Hull Fellowship at Dartmouth College.  

%%%%%%%%%%%%%%%%%%%%%%%%%%%%%%%%%%%%%%%%%%%%%%%%%%%%%%%%%%
\appendix 
\section{CAMB implementation}
\label{appendix}

Here we provide useful notes for implementing the Yang-Mills perturbations in CAMB \cite{Lewis:1999bs}. We will first outline the general approach to implement this model in CAMB and then mention some details in the subsections. Our universal approach is to compute the evolution of the background field $\phi(\tau)$ and then the perturbation equations on top of these. The background field follows a simple ordinary differential equation (ODE) (Eqn.~(\ref{egn:bgeom})), whereas the perturbations technically have to be solved as a system of PDEs due to their spatial variation.

We solve Eqn.~(\ref{egn:bgeom}) with a fourth-order-Runge-Kutta method and store the results in an array. This happens in the \texttt{subroutine init\_background} before CAMB executes the actual Boltzmann integration. To put this model into the code we need to use the right units. CAMB is written in inverse Mpc units, and all energy densities are multiplied by $8 \pi G$. Therefore we define the background field in these units in order to not have to convert it to CAMB units everywhere, $\phi_C$, and the subscript stands for CAMB. We define
\be
	\begin{aligned}
	&\phi'_C = \frac{\phi'}{M_P} \\
	&\phi_C =  \frac{\phi}{\sqrt{M_P}}
	\end{aligned}
	\label{eqn:cambconversion}
\ee
where $\phi'_C$ and $\phi_C$ have mass units of 1 and 1/2 respectively. Physically this is a unusual redefinition, but numerically it makes the equations well behaved as we treat $\phi$ and $\phi'$ as two distinct quantities. With this mapping there is only one scale involved in the equations. $M_P$ always refers to the reduced Planck mass $M_P^{-2} = 8\pi G$. 

There is a natural choice for initial conditions for the background field $\phi$. The energy density (Eqn.~(\ref{eqn:rhop})) splits into the `electric' ($\propto \phi'$) and `magnetic' ($\propto g_\ym \phi^2$) parts. Therefore, we introduce an angle $\theta$ that captures the distribution of energy between these two parts. Energy density considerations can be used as initial conditions for $\phi$ and $\phi'$ by simply assuming a fixed fraction of the YM energy density to the relativistic energy density. Inverting the relationship in Eqn.~(\ref{eqn:rhop}) gives the initial conditions (equivalent to Eqn.~\ref{eqn:initial}),
\be
	\begin{aligned}
		&\phi'_{C, i} = H_0 \sqrt{\frac{2}{3} R_\ym \Omega_{\gamma, 0} \sin^2\theta} \\
		&\phi_{C, i}^2 =\frac{H_0}{g_\ym} \sqrt{\frac{2}{3} R_\ym \Omega_{\gamma, 0} \cos^2\theta}
	\end{aligned}
	\label{eqn:initbg}
\ee
where we choose $g_\ym$, $R_\ym$ and $\theta \in [0, \pi/2]$, which controls the initial relative contributions of $\phi$ and $\phi'$, the electric and magnetic contributions to the YM energy density. Here $\Omega_{\gamma, 0}$ is the radiation energy density fraction today. For our analysis we choose $\theta = \pi/4$ to get a perturbation history close to vanilla $\Lambda$CDM. Deviating from this approach gives more exotic models that we explored around Fig.~\ref{fig:BBvarytheta}. These initial conditions are called once before solving the background equation of motion. 

As shown in Sec.~\ref{sec:perturbations} to first-order the YM fluid behaves like radiation. For the energy density computations in \texttt{equations.f90} we can simply add a routine that multiplies the total radiation density by a factor of $(1+R_\ym)$. The computation of the dark energy density has to be adjusted accordingly for the variable \texttt{P\%omegav} in \texttt{inidriver.f90}.

The Fourier transforms of the perturbation quantities are $k$ dependent and should most efficiently be solved together with the vanilla \lcdm perturbations in the respective routines: The ODE solver for CAMB is \texttt{dverk} which solves first-order ODEs. The second-order perturbation equations therefore have to be split into two first-order equations. The values of the functions and their derivatives get passed on to \texttt{dverk} by the \texttt{subroutine}s \texttt{derivs} for the scalar and \texttt{derivst} for the tensor perturbations. This is where we modify the existing propagation equations for the gravitational waves and add new expressions for the gauge field tensor and scalar perturbations. In the following sections, we will describe in some detail the implementation of this model in CAMB. The tensor modifications are more involved and have a bigger impact on the physics, and hence we will start with them.

\subsection{Tensors}
\label{sec:CAMBtensors}
This subsection describes the implementation of tensor perturbation equations in CAMB. We will outline the general approach computing the ${C_l}$'s via the Boltzmann integration and present the corresponding steps in CAMB.

To obtain the temperature or polarization anisotropy for a given angular mode, one has to integrate the appropriate transfer functions against the initial power perturbations from radiation domination onward until today. See Eqn.~(9) of \cite{Seljak:1996is}, using the appropriate initial conditions. The transfer functions are obtained by integrating the appropriate (tensor) source function against the tensor spherical eigenfunctions (Eqn.~(15) of \cite{Seljak:1996is}). We need to alter the way the source function gets computed, as this contains all the underlying physics.

Generally, the tensor perturbation quantities that enter the source function get defined in \texttt{subroutine derivst} and propagated in time in \texttt{subroutine outputt}. The routine \texttt{derivst} takes first-order ODEs and passes them on to the ODE solver used in CAMB, \texttt{dverk}. We add the new physics to \texttt{derivst} and pass the results on to the higher level ones. 

We extend the number of equations getting propagated by 6, 2 for the left handed gravitational waves (note, we turn one second-order ODE into two first-order ODEs), and 4 for the right- and left handed gauge field tensor perturbations $t_{\rm L/R}$. The variables getting passed on to \texttt{dverk} are labelled \texttt{yt(1), yt(2), ..., yt(n)} and \texttt{ytprime(1), ..., ytprime(n)} for their derivatives. The corresponding physical quantities get mapped as described in Eqns.~(\ref{eqn:cambmap}). Here, \texttt{n} is the variable that has to be increased by 6 in \texttt{subroutine SetupTensorArrayIndices}. We are using the following prescription: in \texttt{dervist} and \texttt{outputt}:
\begin{tabular}{c c c}
	\minipage{0.25\textwidth}
		\bes
			\begin{gathered}
				h_L'' = -k Y_3' \\
				h_L' = -k \, \texttt{shear} = -k Y_3 \\
				h_L = \texttt{Hchi} = Y_2 \\
				Y_2' = -k Y_3 \\ \\
				T_L'' = k Y_n' \\
				T_L' = k Y_n \\
				T_L = Y_{n-1} \\
				k Y_n = Y_{n-1}' \\ \\
		  	\end{gathered}
		\ees
	\endminipage
	\minipage{0.25\textwidth}
		\be
			\begin{gathered}
				h_R'' = -k Y_{n-4}' \\
				h_R' = -k Y_{n-4} \\
				h_R = Y_{n-5} \\
				Y_{n-5}' = -k Y_{n-4} \\ \\
				T_R'' = k Y_{n-2}' \\
				T_R' = k Y_{n-2} \\
				T_R = Y_{n-3} \\
				k Y_{n-2} = Y_{n-3}' \\ \\
		  	\end{gathered}
			\label{eqn:cambmap}
		\ee
	\endminipage
\end{tabular}
where we define
\bes
	T_{L/R} = \sqrt{2} \, a \, t_{L/R}
\ees
to simplify the tensor propagation equation. Also, $Y_n'$ refers to \texttt{ytprime(n)} etc. 

In the following we will translate the perturbation equations into the language of CAMB. A dictionary for the tensor perturbations is given in Table \ref{tab:ScalTranslations} where we collect the most important quantities and procedures to identify the physics in the code more easily. We focus on the files \texttt{cmbmain.f90} and \texttt{equations.f90} which contains most of the physics involved in the Boltzmann integration.
Using the CAMB expressions, the left handed tensor perturbation equation (\ref{eqn:Htens}) for $h_R$ becomes
\bes
	\begin{aligned}
		&h_R''+2\frac{a'}{a}h_R'+k^2 h_R + \frac{2}{a^2 M_P^2}\left({g_\ym}^2 \phi^4-{\phi'}^2\right)h_R \\
		&\quad+ \frac{2\sqrt{2}}{a^2 M_P^2}\phi' T_R' - \frac{2\sqrt{2}}{a^2 M_P^2}{g_\ym}^2 \phi^3 T_R \,\textcolor{red}{+}\, \frac{2\sqrt{2}}{a^2 M_P^2}g_\ym \phi^2kT_R \\
		&\quad = 0 \\
		&\hspace{3.5cm} \Updownarrow \\
		&Y_{n-2} ' = -2\frac{a'}{a} \texttt{shear} + \texttt{Hchi}\left(k+\frac{2}{a^2 k}\left({g_\ym}^2 \phi_C^4 - {\phi'_C}^2\right)\right)  \\
		&\quad + \frac{2\sqrt{2}}{a^2}\phi'_C Y_{n-2} - 2\sqrt{\frac{2 M_P}{a^4}}Y_{n-3} \left(\frac{{g_\ym}^2}{k}\phi_C^3 \,\textcolor{red}{-}\, \frac{g_\ym}{\sqrt{M_P}}\phi_C^2\right)
	\end{aligned}
\ees
where the operations written in \textcolor{red}{red} are changing sign going from right- to left handed propagation. Next, the gauge field tensor perturbation $t$ in of Eqn.~(\ref{eqn:Htens}) becomes, using the background propagation equation (\ref{egn:bgeom}) to eliminate the $\phi'_C$ term
\be
	\begin{aligned}
	Y_{n-2} ' = &Y_{n-3}\left(\textcolor{red}{+}\,2g_\ym \sqrt{M_P} \phi_C - k\right) \\
	&-\sqrt{2}\, \texttt{Hchi}\left(\sqrt{M_P}\frac{{g_\ym}^2}{k}\phi_C^3 \,\textcolor{red}{+}\,g_\ym \phi_C^2 \right) \\
	&-\sqrt{2}\,\phi'_C\,\texttt{shear}
	\end{aligned}
	\label{eqn:Ynprime}
\ee
where again the red operations change sign upon switching from left- to right-hand polarized gravitational waves.

For the perturbations, the initial conditions for the integration of the transfer functions against the initial power spectrum are defined in \texttt{subroutine initialt}. In vanilla CAMB only one tensor perturbation gets propagated. Hence we copy the initial conditions for the second gravitational wave history.

%%%%%%%%%%%%%%%%%%%%%%%%%%%%%%%%%%%%%%%%%%%%%%%%%%%%%%%%%%
\subsection{Scalars}
\label{sec:CAMBscalars}

Here, we will collect results about the scalar implementation in CAMB. See Table \ref{tab:ScalTranslations} for the definitions of the scalar variables in the code. The general strategy we will follow is analogous to the tensors:  solving the two equations of motion for $\{\delta m, \delta \phi \}$ simultaneously. We will use the two constraint equations (Eqns.~(\ref{eqn:scalconstr})) to solve for $\{y, y' \}$.
\begin{table*}
\ra{1.3}
	  \begin{tabular}{l l}
	    \hline
	     	Theory & Code \\ 
		\hline
		\multicolumn{2}{ c }{\texttt{\texttt{cmbmain.f90} }} \\
		\hline
	    	$C_l^{(X)}=(4\pi)^2\int k^2 dk P_h (k) |\Delta_{T,l}^{(X)} (k)|^2$\quad (Eqn.~(9) of \cite{Seljak:1996is}) & \texttt{CalcScalCls} \\
	   	$\Delta^X_{P, l}(k)$ computation & \texttt{\%Delta\_p\_l\_k} \\ 
		$\Delta^{(X)}_{(T,P)l}=\int_0^{\tau_0}d\tau S^{(X)}_{T,P}(k,\tau)\chi^l_k(\tau_0 - \tau)$ \quad (Eqn.~(15) in \cite{Seljak:1996is})  & \texttt{DoFlatIntegration} \\
		\hline
		\multicolumn{2}{ c }{\texttt{equations.f90 ouput(t)} and \texttt{derivs(t)} subroutines variables} \\
		\hline
	    	Source computation $\Delta^{(X)}_{(T,P)l}$& \texttt{output(t)}\\ 
		Eqn.~(29) \& (30) of \cite{cambnotes}: $\pi_\gamma$ \& $E_2$ & \texttt{GaugeInterfaceEvolveTens}: \texttt{y(EV\%g\_ix+2)} \& \texttt{y(EV\%E\_ix+2)}\\
		Eqn.~(29) of \cite{cambnotes}: $\pi_\gamma$  & \texttt{outputt}: \texttt{pig} for tight coupling\\
		$\tau_c^{-1}$ & \texttt{opacity} \\
		$\frac{8}{15}\mathcal{E}_k^{(2)}+\frac{1}{10}I_k^{(2)}$ in \cite{Challinor:2000as} & \texttt{polter} \\
		$I_k^{(2)}$ \cite{Challinor:2000as} & \texttt{pig} \\
		Source $\pi$ in $h''+2\frac{a'}{a}h'+k^2 h = \pi$ & \texttt{rhopi} \\
		$\theta / k$  from Eqn.~(\ref{eqn:prespert}) and \cite{Ma:1995ey} & \texttt{vq} \\
		$\theta'/k$ & \texttt{ayprime(EV\%w\_ix+1)} \\
		$h'/2k$ & \texttt{z} \\
		density contrast $\delta$ & \texttt{clxq} \\
		$\delta'$ & \texttt{ayprime(EV\%w\_ix)} \\
		$\eta k$ metric perturbation from \cite{Ma:1995ey} & \texttt{etak} \\
		$c_s^2$ & \texttt{cs2} \\
		$\pi_\gamma = \frac{32}{45}k\tau_c (v_b + \sigma )$ (Eqn.~(40) in \cite{cambnotes}) & \texttt{pig = 32.\_dl/45/opacity*k*(sigma+vb)} \\
	    	 $M_P^{-2} a^2 \sum_i \rho_i q_i = M_P^{-2} a^2 \sum_i (\rho_i + p_i)v_i = 2k \eta '$  & \texttt{dgq}: total heat flux \\
		$M_P^{-2} a^2 \sum_i \rho_i \delta_i$ & Total matter perturbation \texttt{dgrho} \\
		$M_P^{-2}a^2 \sum_i \rho_i  \sigma_i$ & Total Ma \& Bertschinger \cite{Ma:1995ey} $\sigma$ \texttt{dgs} \\
		$k\eta '$ & \texttt{ayprime(2)} \\
		$\frac{1}{2k}\left(h' + 6 \eta'\right)$ & \texttt{sigma} \\
		$\frac{1}{2k}\left(h'' + 6 \eta''\right)$ & \texttt{sigmadot = -2*adotoa*sigma-dgs/k+etak} \\
		$\eta '$ & $\frac{k}{3}$\texttt{(sigma - z)} \\
		\hline
	  \end{tabular}
	  \caption{Scalar perturbations: translations from theory to code.}
	  \label{tab:ScalTranslations}
\end{table*}
The scalar perturbation equations that are going to source the anisotropic stress (Eqn.~(\ref{eqn:fluidvars})) have to be translated to the synchronous gauge \cite{Ma:1995ey} as do the metric perturbation quantities $b$ and $\Phi_G$ from Eqns.~(\ref{eqn:perts}) to be used inside CAMB. The expressions yield
\bes
	\begin{aligned}
		b &=\alpha - \frac{\eta}{\mathcal{H}} \\
		\Phi_G &= \eta-\frac{\mathcal{H}'}{\mathcal{H}^2}\eta - \frac{\eta '}{\mathcal{H}}-2\mathcal{H}\alpha
	\end{aligned}
\ees
where $\alpha = \frac{1}{2k^2}\left( h' + 6\eta' \right)$. In the calculations we need the first derivative of $\Phi_G$. Computing this is intricate because it requires us to compute $a'''$ and $\eta ''$ which is not directly included in CAMB. However, this is actually not exotic: a description in $a'''$ can be translated into one with the derivative of the equation-of-state parameter $w'$ which is standard in dynamical dark energy models. The $\tau$ derivative of $\Phi_G$ yields, in CAMB variables,
\bes
	\begin{aligned}
	\Phi_G' &= -\frac{a a''' \eta}{a'^2}+\frac{2 a a''^2 \eta}{a'^3}-\frac{a'' \eta}{a'}-\frac{2 \texttt{sigma} a''}{k a} \\
	&+\frac{h'' a}{6 a'}  +\frac{2 \texttt{sigma} a'^2}{k a^2}-\frac{2 \texttt{sigmadot} a'}{k a} \\
	&-\frac{k \texttt{sigmadot} a}{3 a'}+\frac{k \texttt{sigma}}{3}-\frac{k z}{3}
	\end{aligned}
\ees
For $\eta ''$ we need $h''$ and the CAMB variable \texttt{sigmadot} (see Table~\ref{tab:ScalTranslations}). We use Eqn.~(21c) from \cite{Ma:1995ey}, which gives
\be
	h'' = -3\frac{a^2}{M_P^2}\delta P -2\mathcal{H}h' +2k^2 \eta
	\label{eqn:hpp}
\ee
where we only need to plug in the expression for $\delta P$ that is internally solved for in \texttt{derivs}
\be
	\delta P = c_s^2 \delta \rho +\frac{\theta \rho}{k^2} \left[3\frac{a'}{a}\left(1+w\right)\left(c_s^2 - w\right) +w'\right]
	\label{eqn:prespert}
\ee
where in our model we always have $\delta P = \delta \rho/3$. Plugging this into the above Eqn.~\ref{eqn:hpp} gives the expression
\bes
	h'' = -\frac{a^2}{M_P^2}\delta \rho -2\mathcal{H}h' +2k^2 \eta
\ees
which can be expressed purely in terms of CAMB variables. The expression for $a'''$ can be obtained by taking a derivative of the second Friedmann equation and plugging the first one back in together with energy conservation. The result reads
\bes
	a''' = \left(\frac{\rho}{M_P^2}\right)^{3/2}\frac{w(3w-1)}{2\sqrt{3}}a^4   .
\ees

These conversions enter the scalar perturbation equations. Also, we will have to convert the background field $\phi$ into the CAMB variables as before.

The equations need to be put into first-order form. We will use the prescription in the following equations to achieve this:
\begin{tabular}{c c c}
	\minipage{0.25\textwidth}
		\bes
			\begin{gathered}
				\delta \phi'' = k Y_n' \\
				\delta \phi' = k Y_n \\
				\delta \phi =  Y_{n-1} \\
				Y_{n-1}' = k Y_n 
		  	\end{gathered}
		\ees
	\endminipage
	\minipage{0.25\textwidth}
		\bes
			\begin{gathered}
				\delta m'' = k Y_{n-2}' \\
				\delta m' = k Y_{n-2} \\
				\delta m =  Y_{n-3} \\
				Y_{n-3}' = k Y_{n-2} .
		  	\end{gathered}
		\ees
	\endminipage\end{tabular}
With these definitions the equations of motion for $\delta \phi$ and $\delta m$ become, where the index n refers to the extended set of equations in $\texttt{subroutine derivs}$.
\bes
	\begin{aligned}
		&Y_{n-2}' = -k Y_{n-1} - k \phi'_C b +k y' \\
		 &Y_n'  = \frac{1}{k} \left(2 {g_\ym}^2 \phi_C^2 (Y_{n-3}-3 Y_{n-1}) - \phi'_C \left(k^2 b+\Phi_G'\right)\right) \\
		 &\qquad + \frac{4 {g_\ym}^2}{k} \sqrt{M_P} \phi_C^3 \Phi_G - k Y_{n-1}
	\end{aligned}
\ees
These equations get supplemented by the two constraint equations that give $y$ and $y'$:
\be
	\begin{aligned}
		& y = \frac{k(Y_{n-2}-Y_n) + \left(2 {g_\ym}^2 \sqrt{M_P} b \phi_C^3 -\Phi_G \phi'_C \right)}{2 {g_\ym}^2 M_P \phi_C^2+k^2} \\
		& y' = \sqrt{M_P}^{-1} \left(\phi_C \left(b'-\Phi_G\right)-2 \frac{\phi'_C}{\phi_C} y\right) \\
		&\qquad +3 \phi'_C b+3 Y_{n-1}-Y_{n-3}
	\end{aligned}
	\label{eqn:constr}
\ee
which, as described above, get used to compute the scalar perturbation quantities $\delta m$ and $\delta \phi$. 

This appendix serves as a guideline to implement models like this one in CAMB without being too detail oriented. Following these descriptions results in relatively fast CAMB runs that take roughly four times the duration of a vanilla run.

%%%%%%%%%%%%%%%%%%%%%%%%%%%%%%%%%%%%%%%%%%%%%%%%%%%%%%%%%%
\section{Analytic solution to the background equation of motion}
\label{sec:app2}

The solution to the background gauge field equation of motion may be expressed in terms of Jacobi elliptic sine functions. Although these functions can be readily looked up in any math reference book, we give some of their properties here. To be succinct, the  differential equation
\begin{equation}
\frac{d^2 f}{dt^2} + 2 f(t)^3 = 0
\label{eqn:master}
\end{equation}
has solution
\begin{equation}
f(t) = c_1 {\rm sn}(c_1 t + c_2 | -\hspace{-0.25em}1)
\label{eqn:soln}
\end{equation}
where the function ${\rm sn}(u|m)$ is the Jacobi elliptic sine-amplitude function. This and related functions are described in Sec. 8.14 of Ref.~\cite{GR}. The applicability of the above solutions to the description of both a  scalar field $\phi$ with a $\lambda \phi^4$ self-interaction and $SU(2)$ Yang-Mills was presented in Ref.~\cite{Actor:1979in}.  
 
\begin{figure}[b] 
	\includegraphics[width=\linewidth]{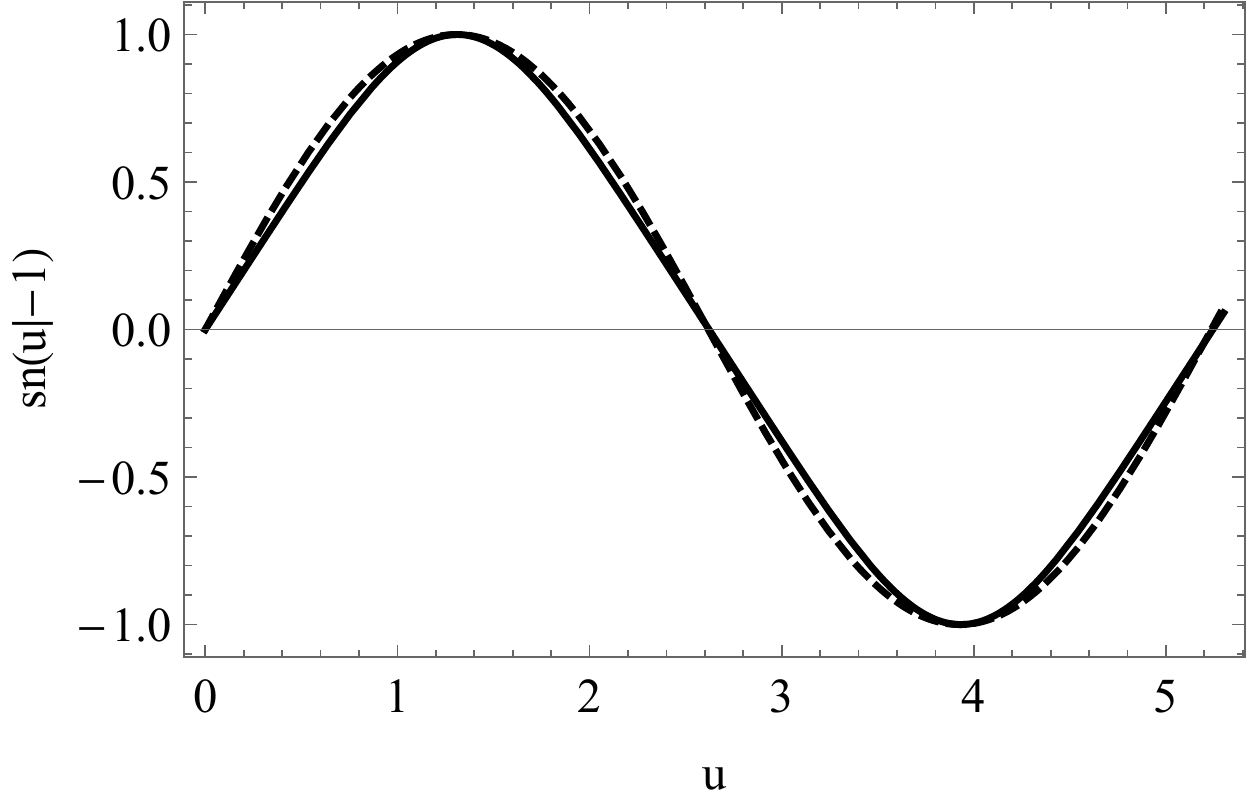}
	\hspace{1.0cm}
	\includegraphics[width=\linewidth]{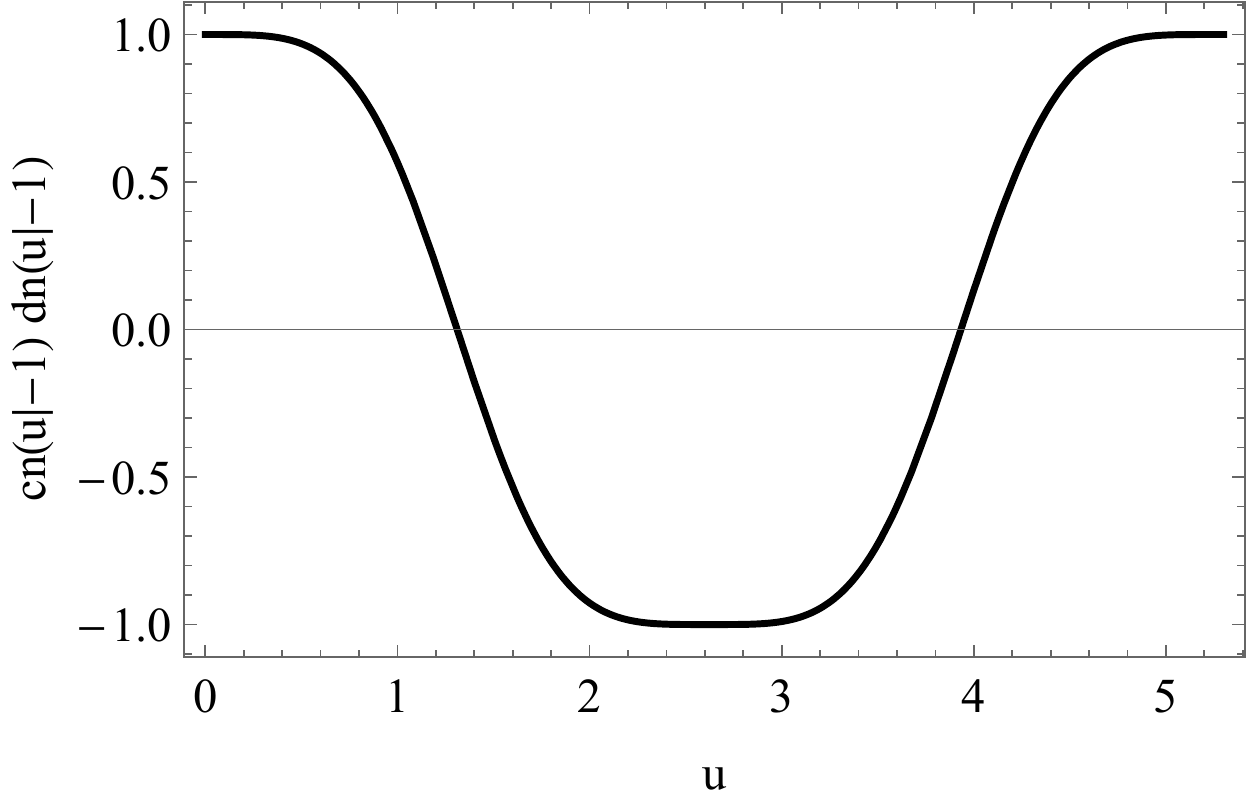}
	\caption{Jacobi elliptic functions ${\rm sn}(u|-\hspace{-0.25em}1)$ and its derivative ${\rm cn}(u|-\hspace{-0.25em}1) \, {\rm dn}(u|-\hspace{-0.25em}1)$ are illustrated. The dashed line in the first panel corresponds to a regular sine function with a scaled argument: $\sin(\frac{6}{5} u)$.}
	\label{fig:jacobi}
\end{figure}

A few useful facts are that ${\rm sn}(u|m)$ can be evaluated by the following recipe:
\begin{equation}
{\rm sn}(u|m) = \sin\phi, \, \quad {\rm for}\, \quad u = \int_0^\phi \frac{d\theta}{\sqrt{1- m \sin^2\theta}}.
\end{equation}
Our notation is consistent with Ref.~\cite{GR} as well as Mathematica \cite{Mathematica10}. (In some notation, $m$ is replaced by $m^2$ in the integral on the right. Elsewhere, it is common to drop the $m$ so that ${\rm sn}(u|m)$ becomes ${\rm sn} \,u$.) The derivative is $d/du \,{\rm sn}(u|m) = {\rm cn}(u|m){\rm dn}(u|m)$, which consists of Jacobi elliptic cosine-amplitude and delta-amplitude functions. Together they satisfy the identity
\begin{equation}
{\rm sn}(u|m)^4 + {\rm cn}(u|m)^2\, {\rm dn}(u|m)^2 = 1.
\end{equation}
We are interested in the case $m=-1$, for which these functions oscillate with period $T = \Gamma(1/4)^2/\sqrt{2 \pi}$. In the high-frequency limit, it may be useful to average over the oscillation period. In this case
\begin{equation}
	\begin{aligned}
		&\langle {\rm sn}(u|-\hspace{-0.25em}1)^2\rangle  = \langle {\rm dn}(u|-\hspace{-0.25em}1)^2\rangle-1 = 1-\langle {\rm cn}(u|-\hspace{-0.25em}1)^2\rangle \\
		&\qquad = \frac{8 \pi^2}{\Gamma(1/4)^4},\\
		 &\langle {\rm sn}(u|-\hspace{-0.25em}1)^4\rangle = \frac{1}{3}.
	\end{aligned}
\end{equation}
Illustrative plots of these functions are provided in Fig.~\ref{fig:jacobi}.

%%%%%%%%%%%%%%%%%%%%%%%%%%%%%%%%%%%%%%%%%%%% 

\end{document}